\documentstyle[12pt,epsf]{article}
\newcommand{\beq}{\begin{equation}}
\newcommand{\eeq}{\end{equation}}
\newcommand{\beqa}{\begin{eqnarray}}
\newcommand{\eeqa}{\end{eqnarray}}
\newcommand{\ba}{\begin{array}}
\newcommand{\ea}{\end{array}}
\newcommand{\CR}{\nonumber \\}

\newcommand{\bra}{\langle}
\newcommand{\ket}{\rangle}

\newcommand{\half}{{1\over 2}}

\newcommand{\wt}{\widetilde}
\newcommand{\wh}{\widehat}

\newcommand{\bA}{{\bf A}}
\newcommand{\bB}{{\bf B}}
\newcommand{\bC}{{\bf C}}
\newcommand{\bD}{{\bf D}}
\newcommand{\bE}{{\bf E}}
\newcommand{\bH}{{\bf H}}
\newcommand{\bJ}{{\bf J}}
\newcommand{\bP}{{\bf P}}
\newcommand{\bQ}{{\bf Q}}

\newcommand{\bX}{{\bf X}}
\newcommand{\bZ}{{\bf Z}}
\newcommand{\bj}{{\bf j}}
\newcommand{\bs}{{\bf s}}
\newcommand{\bz}{{\bf z}}

\newcommand{\cN}{{\cal N}}
\def\br#1#2{\bX_{\bf [#1,#2]}}

\def\K#1#2{K_{[#1,#2]}}

\def\u1#1{{\langle #1 \rangle}}
\def\bfgr#1{\mbox{\boldmath $\bf #1$}}

\begin{document}

\begin{titlepage}
\begin{flushright}
{\tt hep-th/9909122} \\
UTHEP-408 \\
September, 1999
\end{flushright}
\vspace{0.5cm}
\begin{center}
{\Large \bf 
Mordell-Weil Lattice via String Junctions
\par}
\lineskip .75em
\vskip2.5cm
\normalsize
{\large Mitsuaki Fukae, Yasuhiko Yamada}
\vskip 1.5em
{\large\it Department of Mathematics, Kobe University \\
Rokko, Kobe 657-8501, Japan}
\vskip1cm
{\large Sung-Kil Yang}
\vskip 1.5em
{\large\it Institute of Physics, University of Tsukuba \\
Ibaraki 305-8571, Japan}
\end{center}
\vskip3cm
\begin{abstract}
We analyze the structure of singularities, Mordell-Weil lattices and torsions
of a rational elliptic surface using string junctions in the background of 12 
7-branes. The classification of the Mordell-Weil lattices due to Oguiso-Shioda
is reproduced in terms of the junction lattice. In this analysis an important 
role played by the global structure of the surface is observed. It is then 
found that the torsions in the Mordell-Weil group are generated by the 
fraction of loop junctions which represent the imaginary roots of the loop 
algebra $\widehat E_9$. From the structure of the Mordell-Weil lattice we 
find 7-brane configurations which support non-BPS junctions carrying
conserved Abelian charges.
\end{abstract}

\end{titlepage}

\tableofcontents

\baselineskip=0.7cm

\section{Introduction}

In our previous paper \cite{YY}, and in a related work \cite{SZ}, elliptic
curves have been constructed for the 7-brane configurations on which the
affine Lie algebras $\wh E_n$ $(1 \leq n\leq 8)$ and $\wh{\wt E}_n$ $(n=0,1)$
are realized. Upon deriving these curves from a rational elliptic surface $S$,
we recognize that the brane picture is very efficient to deal with the
geometry of $S$. In this construction, however, we have only probed the 
{\it local} geometry of $S$ with the aid of the 7-brane technology. Our purpose
in this paper is to show that 7-branes and string junctions stretched among
them precisely capture the {\it global} structure of a rational elliptic
surface.

To explain what kind of global structures we will discuss, let us start with
briefly reviewing the heterotic string/F-theory duality. 
Duality between F-theory on an 
elliptic $K3$ surface $S_F$ and the heterotic string theory 
on a two-torus $T^2$ has been the source of 
inspiring various duality relations in lower dimensions \cite{Vafa,Sen-1,MV-1}.
In the type IIB picture, singularities of $S_F$ are described in terms of
coalescing 7-branes which are in general mutually non-local. The sub-lattice
$\Gamma^{18,2}$ of the 22-dimensional homology lattice $\Gamma^{19,3}$
of $K3$ is spanned 
by junctions stretched among 7-branes \cite{DHIZ-3}. When the size
of $T^2$ is very large, $S_F$ can be viewed as consisting of two rational
elliptic surfaces $S_1$, $S_2$, each of which may be associated to one of the
heterotic $E_8 \times E_8$ gauge groups \cite{FMW}. 
More precisely, in this regime,
$S_1$ and $S_2$ intersect along an elliptic curve $E_*$ whose complex structure
is identified as that of $T^2$ on the heterotic side. While keeping $E_*$
fixed, then, deformations of the complex structure of $S_i$ are dual to
deformations of the corresponding $E_8$ bundle on $T^2$ \cite{FMW,AM-1}.

The gauge symmetry in the heterotic string
is now understood in terms of shrinking two-cycles in $S_i$, or equivalently 
coalescing 7-branes, on the F-theory side.
As found in \cite{Asp,AM-2}, while the resulting singularity fixes the root 
lattice, and hence the gauge symmetry {\it algebra}, the torsion part of the
Mordell-Weil group plays a crucial role to determine the gauge {\it group}.
The Mordell-Weil group, as will be described in more detail in the text, is
an Abelian group generated by rational sections of the elliptic fibration,
and decomposed into its free part and its torsion part. 
When the torsion part is non-trivial
the corresponding gauge group acquires non-trivial $\pi_1$, and thus 
there appears non-simply-connected gauge groups \cite{AM-2}.

Thus it is seen that the Mordell-Weil group reflects a global structure of
a rational elliptic surface. Furthermore it is known that
the Mordell-Weil group equipped
with a natural bilinear pairing possesses the lattice structure which is
referred to as the Mordell-Weil lattice \cite{Shio}. The classification of
the Mordell-Weil lattice of a rational elliptic surface has already been 
completed, thanks to Oguiso-Shioda \cite{OS}.

In this paper, describing 
the singularity structures of a rational elliptic surface $S$ 
in terms of 12 7-branes with trivial total monodromy,
we will show that the junction lattices on the 7-brane backgrounds precisely 
produce all the Mordell-Weil lattices of $S$ as listed in \cite{OS}.
Especially the torsion elements are found to be identified as the ``fraction''
of global loop junctions which, on the other hand, are related to the
imaginary roots of $E$-type affine Lie algebras. 

This paper is organized as follows:
In section 2, we introduce some elementary 
arithmetic of elliptic curves and the notion of the Mordell-Weil group 
of $S$ using several explicit examples of $S$ presented as 
elliptic curves.
In section 3, after a short review of 7-branes and junctions,
we determine the brane configurations which describe the structure
of the Mordell-Weil lattice given in \cite{OS}. In this calculation
we recognize a non-trivial role played by the global structure of the surface.
In section 4, the charge integrality condition is re-considered
to obtain the weight lattice and torsions from string junctions.
In particular, torsions are expressed as fractional loop junctions.
Finally we discuss possibly stable non-BPS states in F-theory
on $K3$ on the basis of the structure of the Mordell-Weil lattice.

\section{Preliminaries}

\renewcommand{\theequation}{2.\arabic{equation}}\setcounter{equation}{0}

\subsection{Mordell-Weil group}

Consider an elliptic curve $E$ defined over $\bQ$
\beq
y^2=4 x^3-g_2 x-g_3, \hskip10mm  g_2,\; g_3 \in \bQ.
\label{Weierstrassform}
\eeq
Any cubic in $\bP^2$ can be transformed into this canonical form. 
A point $P=(x,y)$ on $E$ is called rational point if
$x,y \in \bQ$, the totality of which is denoted as $E(\bQ)$.
Determination of the structure of $E(\bQ)$ is a deep arithmetic problem.
The fundamental fact is that $E(\bQ)$ has a structure of
an Abelian group, known as the Mordell-Weil group \cite{ST}.

In order to see the group law explicitly, let us recall that 
a point $P=(x,y)$ on $E$ is parametrized by the Weierstrass $\wp$-function
as $x=\wp(t)$, $y=\wp'(t)$. 
The addition formulae for $\wp$ read
\beqa
&&\wp(s+t)=-\wp(s)-\wp(t)
+\frac{1}{4}\left(
\frac{\wp'(s)-\wp'(t)}{\wp(s)-\wp(t)}
\right)^2, \CR
&&\wp'(s+t)=-\wp'(s)-\wp'(t)
+\frac{1}{4}\left(
\frac{\wp'(s)-\wp'(t)}{\wp(s)-\wp(t)}
\right)^3+3 
\frac{\wp(s)\wp'(t)-\wp(t)\wp'(s)}{\wp(s)-\wp(t)}.
\label{peadd}
\eeqa
It is obvious that, given $P_i=(x_i,y_i)=(\wp(t_i),\wp'(t_i)) \in E(\bQ)$
with $i=1,2$, the ``sum'' $P_3=P_1+P_2$
defined by $P_3=(x_3,y_3)=(\wp(t_1+t_2),\wp'(t_1+t_2))$ is also in $E(\bQ)$. 
The special point at infinity $O=(\infty , \infty)$ is regarded as the
unit of this addition. The inverse of $P=(x,y)$ is given by $-P=(x,-y)$.

When the curve $E$ is expressed as a complex torus
$E_{\tau}=\bC/(\bZ+\tau \bZ)$, the addition is nothing but the
usual sum of complex numbers $t_i \in E_{\tau}$.
Note that if $t_1+t_2+t_3=0$ in $E_{\tau}$ then
\beq
\det
\left|
\begin{array}{ccc}
\wp(t_1)&\wp'(t_1)&1 \\
\wp(t_2)&\wp'(t_2)&1 \\
\wp(t_3)&\wp'(t_3)&1 
\end{array}
\right|=0.
\eeq
This means that three distinct points
$P_i \in E$ satisfy $P_1+P_2+P_3=0$ if and only if
they are on the same line.
This geometric rule of the addition, classically known to
Fermat and Euler, is applicable 
for any cubic which is not necessarily in the canonical form
(\ref{Weierstrassform}).

Mordell's theorem says that the group $E(\bQ)$ is
finitely generated, 
\beq
E(\bQ) \equiv \bZ^{\oplus r} \oplus E(\bQ)_{\rm{tor}},
\eeq 
where the torsion part $E(\bQ)_{\rm{tor}}$ is generated by
(at most two) generators $P \in E(\bQ)$ such that
$m P=O$ for some $m \in \bZ_{>0}$ (see \cite{AM-2} and references therein).

\subsection{Elliptic surfaces}

What we have described in the previous subsection can be readily generalized
for other field $K$ than $\bQ$.
The relevant case for our analysis in this paper is that $K$ is
a field of rational functions of one variable, namely
$K=\bC(z)=\{ \ a(z)/b(z) \ \vert \ a(z), b(z)$ polynomials in $z \}.$
Then the curve
\beq
y^2=x^3+f(z) x+g(z),
\label{fgcurve}
\eeq
where $f(z),\, g(z) \in K$, represents an elliptic surface 
$S$ over $\bP^1$ which is a family of elliptic curves 
$E=\{ \, (x,y) \, \}$ parametrized by $z \in \bP^1$.
Every elliptic surface over $\bP^1$ with section can be recast
in this form. At the point $z$ on $\bP^1$ where the discriminant 
$\Delta=4 f(z)^3+27 g(z)^2$ vanishes,
the fiber becomes singular. Possible singular fibers were classified by
Kodaira, according to which the blowing-up diagram of each singularity is
represented by the Dynkin diagram of $A$, $D$ and $E$ type \cite{Kod}.
In the IIB brane picture, the singularities correspond to
the coinciding 7-branes \cite{Vafa}.

Now, a point $P$ in the Mordell-Weil group $E(K)$ is a solution
$P=(x(z),y(z))$ of (\ref{fgcurve}) such that $x(z),\, y(z) \in K$.
In view of the elliptic surface $S$, a point 
$P \in E(K)$ is a rational section
of the elliptic fibration. Applying the addition formulae (\ref{peadd}) 
fiberwise, we see that $E(K)$ has a structure of an Abelian group.

In the following let us take several examples of elliptic surfaces and
compute rational sections to show the Abelian group property explicitly.
We note in passing that determining sections of elliptic surfaces is essential
when one constructs the Seiberg-Witten differential in $\cN =2$ supersymmetric
gauge theories with matters \cite{SW-2,MN,NTY}. In each example below,
No.\,$\sharp$ refers to the entry $\sharp$ of Table \ref{config} which will
appear in section 3.

\noindent {\bf Example}. (No.72) : $\cN =2$ $SU(2)$ theory with 
massless $N_f=3$ flavors \cite{SW-2}
\beq
y^2=x^2(x-z)-\frac{\Lambda_3^2}{64} (x-z)^2.
\eeq
The discriminant is $\Delta=\Lambda_3^2 (256 z-\Lambda_3^2) z^4 /4096$
and the singularities are given by
\beq
A_3 : (z=0), \hskip10mm 
D_5 : (z=\infty), \hskip10mm
A_0 : (z=\Lambda_3^2/256).
\eeq
This curve has four sections generated by a single element $P$,
\beqa
&&P=\left(0,\;-\frac{i \Lambda_3}{8} z \right), 
\hskip10mm
2 P=\left(z,\; 0 \right), \CR
&&3 P= \left(0,\; \frac{i \Lambda_3}{8} z \right), 
\hskip10mm
4 P=O=(\infty,\infty).
\eeqa
Hence the Mordell-Weil group consists of the torsion part only, {\it i.e.}
$E(K)=\bZ/4 \bZ$.

\noindent {\bf Example}. (No.66) : Massless $\widehat E_3$ 
curve \cite{MNW,YY,SZ}
\beq
y^2=x^3+(z^2+10 z-23) x^2+128 (1-z) x.
\eeq
The discriminant is
$\Delta=-16384(z+1)^3(z-1)^2(z+17)$ and the singularities are
\beq
A_5 : (z=\infty), \hskip10mm
A_2 : (z=-1), \hskip10mm
A_1 : (z=1), \hskip10mm
A_0 : (z=-17).
\eeq
This curve has six sections generated by a single element $P$,
\beqa
&& P=(8(1-z),\; -8(1-z^2)), \hskip10mm
2 P=(16,\; -16(1+z)), \CR
&& 3 P=(0,\; 0), \hskip37mm
4 P=(16,\; 16(1+z)), \CR
&& 5 P=(8(1-z),\; 8(1-z^2)), \hskip10mm
6 P=O=(\infty,\infty). 
\eeqa
Hence we have $E(K)=\bZ/6 \bZ$.

\noindent {\bf Example}. (No.44 and its degeneration to No.70) : Massive and 
massless $\widehat E_1$ curve \cite{MNW,YY,SZ} \\
The massive $\widehat E_1$ curve reads
\beq
y^2=x^3+\left(z^2-2 (p+\frac{16}{p}) z+p^2-224
\right) x^2+\frac{65536}{p^2} x.
\eeq
The discriminant is
\beq
\Delta=-4294967296 (z-p-16)(z-p+16)(p z-p^2+16 p -32)
(p z-p^2-16 p -32)/p^6.
\eeq
The singularities are
\beqa
A_7 &:& (z=\infty), \CR
A_0+A_0+A_0+A_0 &:& (z=p \pm 16, \ p+\frac{32}{p} \pm 16).
\eeqa
The Mordell-Weil group $E(K)$ is generated by
\beqa
&& P=(0,\; 0), \hskip10mm
Q=\left(256,\; 256 z-256\frac{16+p^2}{p} \right), \CR
&& R=\left(\frac{256}{p^2},\; 
\frac{256}{p^2} z-256\frac{16+p^2}{p^3} \right) .
\eeqa
Since $2 P=O$, $Q+R=P$ we have
$E(K)=\bZ \oplus \bZ/2 \bZ$
where the free part $\bZ$ and the torsion part $\bZ/2 \bZ$ are
generated by $Q$ and $P$, respectively.

Under the degeneration at $p=1$ (massless $\widehat E_1$ curve) 
the discriminant becomes
\beq
\Delta=-4294967296(z+15)(z-49)(z-17)^2
\eeq
and the singularities are
\beq
A_7 : (z=\infty ), \hskip10mm
A_1 : (z=17), \hskip10mm
A_0+ A_0 : (z=-15,49).
\eeq
In this case, we have $Q=R$ and $4 Q=2 P=O$,
thereby $E(K)$ reduces to $E(K)=\bZ/4 \bZ$.

\noindent {\bf Example}. (No.27) : Massive $A_2$ curve \cite{NTY}
\beq
y^2=x^3+u x+v+z^2.
\label{A2curve}
\eeq
The discriminant is $\Delta=27 z^4+54 v z^2+4 u^3+27 v^2$ and
the singularities are
\beqa
E_6 &:& (z=\infty), \CR
A_0+A_0+A_0+A_0 &:& (\Delta(z)=0),
\eeqa
This curve has six fundamental sections,
\beq
\pm P_{i}=(a_i,\; \pm z), \quad i=1,2,3,
\eeq
where $a_i$ are determined through $(x-a_1)(x-a_2)(x-a_3)=x^3+u x+v$.
These six sections $\pm P_{i}$ correspond to the six fundamental weights of
the $A_2$ algebra $\pm \Lambda_1$, $\pm (\Lambda_2-\Lambda_1)$
and $\pm(-\Lambda_2)$.
By addition formulae (\ref{peadd}), one can generate a section $P_{l,m}$
corresponding to a weight $\lambda=l \Lambda_1+m \Lambda_2$
for each $(l,m) \in \bZ^{\oplus 2}$.
Thus we see that $E(K) \equiv \bZ^{\oplus 2}$ as an abstract group, but
it has a more detailed structure as a lattice $E(K)=A_2^{\ast}$.
Here a ``lattice'' is introduced as a free Abelian group $L=\bZ^{\oplus r}$ 
with a symmetric bilinear form $(\ , \ ) : L \otimes L \rightarrow \bZ$.
The equality $E(K)=A_2^{\ast}$ means the isomorphism as a lattice.
The bilinear form on $E(K)$ is defined by the ``height pairing''
which can be explicitly evaluated in terms of an intersection pairing
on an elliptic surface $S$, see next section.

Finally it is mentioned that, extending the computation in \cite{NTY}, we have 
shown that the Seiberg-Witten differential for the curve (\ref{A2curve})
can be constructed so that it has the poles with residues located on
sections (\ref{A2section}). The details may appear elsewhere.

\section{Mordell-Weil lattice vs. junction lattice}

\renewcommand{\theequation}{3.\arabic{equation}}\setcounter{equation}{0}

\subsection{Mordell-Weil lattice}

Let us start with recapitulating what we have discussed in the previous
section, then we introduce the notion of the Mordell-Weil lattice,
following \cite{Shio}.
We take an elliptic curve $E$ defined over a field $K$
\beq
y^2=x^3+f x+g, \hskip10mm  f,\; g \in K.
\label{cubic}
\eeq
A point $P=(x,y) \in K^2$ on the curve $E$ is called 
$K$-rational point. Let $E(K)$ denote a set of all the $K$-rational points
plus the point at infinity $O=(\infty ,\infty)$.
For given $P_i=(x_i,y_i) \in E(K)$ with  $i=1,2$, 
the third point $P_3=P_1+P_2=(x_3,y_3)\in E(K)$ is defined by
\beq
x_3=-x_1-x_2+m^2, \hskip10mm
y_3=-m(x_3-x_1)-y_1, 
\label{addition}
\eeq
where
\beq
m= \left\{
\begin{array}{ll}
(y_1-y_2)/(x_1-x_2), & \quad \mbox{if $P_1 \neq P_2$}, \\
(3 x_1^2+f)/(2 y_1), & \quad \mbox{if $P_1=P_2$}.
\end{array}\right.
\eeq
With respect to this addition law, $E(K)$ has a structure of Abelian group,
called the Mordell-Weil group.
The point at infinity $O \in E(K)$ is the unit of this addition. 

In this paper we are mainly concerned with the case of $K=\bC(z)$ 
(the field of rational functions on $z$). Then (\ref{cubic}) is naturally 
considered as an elliptic surface $S$ over $\bP^1$, for which
the points $P=(x(z),y(z)) \in E(K)$ represent
the rational sections of this fibration. The Kodaira classification of 
singular fibers is presented in Table \ref{kunihiko}.
We denote by $T$ the lattice corresponding to the singular fibers.

\renewcommand{\arraystretch}{1.2}
\begin{table}
\begin{center}
\begin{tabular}{||c|c|c||} \hline
fiber type & singularity lattice & 7-branes \\
\hline
${\rm I}_n \ (n \geq 1)$ & 
$A_{n-1}$ & $\bA^n=\bA_{\bf n-1}$ \\
${\rm II,III,IV} \ (n=0,1,2)$  &  
$A_n$ & $\bA^{n+1} \bC=\bH_{\bf n}$ \\
${\rm I}_n^* \ (n \geq 0)$ & 
$D_{n+4}$ & $\bA^{n+4}\bB\bC=\bD_{\bf n+4}$ \\
${\rm II}^*,{\rm III}^*,{\rm IV}^*  \ (n=8,7,6)$ &  
$E_n$ & $\bA^{n-1}\bB\bC^2=\bE_{\bf n}$ \\
\hline
\end{tabular}
\end{center}
\caption{Kodaira classification, ADE singularities and 7-branes.}
\label{kunihiko}
\end{table}

There exists a deep relation between the singularities $T$ and the
sections $E(K)$, reflecting the global structure of the elliptic surface $S$.
The essential idea in studying such relation
is to equip $E(K)$ with the lattice structure which
can be described in terms of intersections on $S$.

The lattice structure, or the height pairing $(\ ,\ )$
of the Mordell-Weil group $E(K)$ was introduced by Shioda
using the intersection pairing \cite{Shio}. The Mordell-Weil group
equipped with the height pairing is called the Mordell-Weil lattice.
In Theorem 8.1 \cite{Shio}, the height pairing
$(P,Q)$ for sections $P,Q \in E(K)$ is explicitly calculated as follows:
\beqa
&&(P,Q)=P\cdot O+Q \cdot O-P \cdot Q+\chi({\cal{O}}_S)-
\sum_v {\rm contr}_v(P,Q),
\CR
&&(P,P)=2 \chi({\cal{O}}_S)+2 P \cdot O-
\sum_v {\rm contr}_v(P),
\label{PPformula}
\eeqa
where $P \cdot Q$ denotes the intersection pairing,
$\chi({\cal{O}}_S)$ is the arithmetic genus of $S$
($=1$ for rational elliptic surfaces)
and ${\rm contr}_v$ is the local contribution from each singular point $v$
\beq
{\rm contr}_v(P,Q)=
\left\{
\begin{array}{cl}
0,               & {\rm if} \ i=0 \ {\rm or} \ j=0, \CR
(C_v^{-1})_{ij}, & {\rm otherwise}
\end{array}
\right.
\eeq
and ${\rm contr}_v(P)={\rm contr}_v(P,P)$.
Here $C_v$ is the Cartan matrix (of finite type) corresponding
to the singularity at $v$ and 
the indices $i,j$ label the components of the singular
fibers with which $P$ or $Q$ intersects.
The component intersecting with the zero section $O$ is specified by $i=0$. 

\noindent
{\bf Example}. (No.27 : The massive $A_2$-curve (\ref{A2curve}) continued)

\noindent
The rational section $P_{l,m}$
corresponding to a weight vector $\lambda=l \Lambda_1+m \Lambda_2$
takes the form
\beq
x(z)=\frac{\phi(z)}{\chi(z)^2}, \hskip10mm
y(z)=\frac{\psi(z)}{\chi(z)^3},
\label{A2section}
\eeq
where
$\deg \psi(z)=d_{l,m}=l^2+l m+m^2$,
$\deg \chi(z)=\frac{1}{3}d_{l,m}-1$ 
[ resp. $ \frac{1}{3}d_{l,m}-\frac{1}{3}$ ] and
$\deg \phi(z)=\frac{2}{3}d_{l,m}$ 
[ resp. $ \frac{2}{3}d_{l,m}-\frac{2}{3}$ ]
for $l \equiv m$ (mod 3) [ resp. otherwise ].
For this section $P_{l,m}$, the second formula in
(\ref{PPformula}) can be checked since we evaluate
$(P_{l,m},P_{l,m})=(l \Lambda_1+m \Lambda_2)^2=\frac{2}{3} d_{l,m}$,
${\rm contr}_{z=\infty} (P_{l,m})=0$ [ resp. $\frac{4}{3}$ ]
and $O \cdot P_{l,m}=\deg \chi(z).$\footnote{
In the homogeneous coordinates $(X:Y:Z)$, the curve and sections
are rewritten as
$Z Y^2=X^3+f(z) X Z^2+g(z) Z^3$,
$P=(X:Y:Z)=(\phi \chi, \psi, \chi^3)$.
Hence, the intersection of $P$ with zero-section $O=(0:1:0)$ is given
by $\deg \chi(z)$.}

Noting that the structure of the Mordell-Weil lattice $E(K)$ is
essentially determined by the singularity lattice $T$, Oguiso-Shioda 
classified the Mordell-Weil lattice $E(K)$ of a rational elliptic 
surface \cite{OS}. This will be described in detail in section 3.3.
Our task is now to figure out how the
Mordell-Weil lattice $E(K)$ and the singularity lattice $T$ are related to
each other in terms of the junctions on  
rational elliptic surfaces.\footnote{
In case of more general elliptic surfaces, elliptic K3 for instance,
the junction lattice will contain transcendental cycles also.}

\subsection{Brane configurations and junction lattice}

We concentrate on a rational elliptic surface $S$ defined by (\ref{cubic})
where $f=f(z)$ and $g=g(z)$ are some polynomials in $z$, and 
$\deg f \leq 4, \ \deg g \leq 6$ with $\Delta \neq {\rm constant}$.
Generically there exist 12 singular points $z_i$ where $\Delta(z_i)=0$. 
Each singular fiber at $z=z_i$ has a local monodromy 
$K_{[p_i,q_i]}$ labeled by $p_i,q_i \in \bZ$. Here the monodromy matrix
takes the form
\beq
\K{p}{q} =\pmatrix{1+pq & -p^2 \cr q^2 & 1-pq \cr},
\eeq
corresponding to a vanishing cycle $p \alpha+q \beta$ with $\alpha , \beta$
being homology cycles of a fiber torus.
Physically a singular point is interpreted as the position
of a 7-brane $\br{p}{q}$ on which a ($p,q$)-string with a boundary
homologous to $p \alpha+q \beta$ can end \cite{Vafa}.
Among various $(p,q)$ 7-branes $\br{p}{q}$ it is convenient to express the
representative ones as $\br{1}{0}=\bA, \br{1}{-1}=\bB, \br{1}{1}=\bC$.
The Kodaira singularities can be then described as a coalescence
of collapsible sub-configurations of 7-branes, 
see Table \ref{kunihiko} \cite{dWZ,DHIZ-2}.

To specify a brane configuration, we place the 12 branes, say, on the real
axis of the $z$-plane and draw downwards the branch cuts emanating from the
branes. Thus, for a brane configuration 
\beq
\br{p_1}{q_1} \br{p_2}{q_2} \cdots \br{p_{12}}{q_{12}},
\eeq
we have the total monodromy
\beq
K=\K{p_{12}}{q_{12}} \cdots \K{p_2}{q_2} \K{p_1}{q_1}
\eeq
which should be trivial, {\it i.e.} $K=1$, 
to describe a rational elliptic surface $S$.
A standard realization of such a brane configuration is \cite{DHIZ-3}
\beq
\widehat{\bE}_{\bf 9}=\bA^8 \bB \bC \bB \bC.
\label{hatE9}
\eeq

A topological configuration of strings (or string junctions) 
associated to the branes can be parameterized as
\beq
\bJ=\sum_{i=1}^{12} Q_i \bs_i \hskip10mm  {\rm or} \hskip10mm 
\bJ=(Q_1,Q_2,\ldots,Q_{12}),
\eeq
where $\bs_i$ stands for the outgoing $(p_i,q_i)$-string starting
at $\br{p_i}{q_i}$ and $Q_i$ is the net number of outgoing $(p_i,q_i)$-strings.
By definition the charges $Q_i$ must be integral and are called the invariant
charges \cite{dWZ}. The total $(p,q)$ charges of a string junction are
\beq
(p,q)=\sum_{i=1}^{12} Q_i (p_i,q_i).
\label{totalpq}
\eeq
A junction represents a closed two-cycle in $S$ if and only if
its total charges vanish. 

\begin{figure}
\hspace{1.5cm}
\epsffile{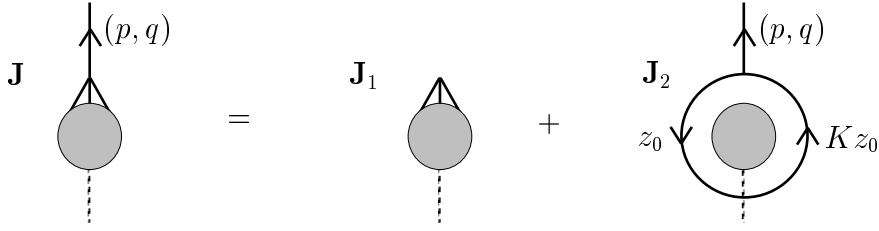}
\caption{A string junction.}
\label{parts}
\end{figure}

A $(p,q)$-junction $\bJ$ is in general decomposed 
into two pieces $\bJ_1,\, \bJ_2$ as depicted in Figure \ref{parts}. 
If $\bJ$ is a singlet with respect to the symmetry
realized on the branes, then $\bJ_1=0$.
The charge ${\bf z}_0=(r,s)$ of the tadpole loop $\bJ_2$ is determined
by charge conservation at the trivalent vertex. When the tadpole loop
goes around the ${\bf A_n}$, ${\bf D_n}$ and ${\bf E_n}$ branes of 
Table \ref{kunihiko}, 
the loop charges $(r,s)$, invariant charges $Q_i$ and
outgoing charges $(p,q)$ are related as follows:
\beqa
\label{Antad}
{\bf A_n} :
&&\bfgr{\delta}_{r,s}=-s(\bs_1+\cdots +\bs_{n+1}), \CR
&&(p,q)=(-(n+1) s,\, 0), \\
\bD_n :
&&\bfgr{\delta}_{r,s}=-s(\bs_1+\cdots +\bs_n)-(r-(n-1)s)\bs_{n+1}
-(r-(n-3)s)\bs_{n+2}, \CR
&&(p,q)=(-2 r+(n-4)s,\, -2 s), \\
\label{Dntad}
{\bf E_n} :
&&\bfgr{\delta}_{r,s}=-s(\bs_1+\cdots +\bs_{n-1})
-(r-(n-2)s)\bs_{n+1}-(r-(n-4)s)(\bs_{n+1}+\bs_{n+2}),\CR
&&(p,q)=(-3 r+(2 n-9)s,\, -r+(n-6)s).
\label{Entad}
\eeqa
It should be remembered that if one employs a different 
7-brane configuration from the ones given in Table \ref{kunihiko} 
to describe the ADE singularities,
the assignment of the $(r,s)$ charges will also change.

The junctions form a lattice which is endowed with a symmetric intersection
pairing defined by \cite{dWZ}
\beqa
&& (\bs_i,\bs_i)=-1, \CR
&& (\bs_i,\bs_j)=(\bs_j,\bs_i)=\frac{1}{2} \left(p_i q_j-p_j q_i \right), \quad
\mbox{for $i < j$}.
\eeqa
For the sub-configurations in Table \ref{kunihiko}, the junction lattice with
$(p,q)=0$ is known to be isomorphic to the corresponding root 
lattice \cite{dWZ}. In these cases, the root junctions represent the
BPS strings responsible for the symmetry enhancement. 

\begin{figure}
\hspace{1.5cm}
\epsffile{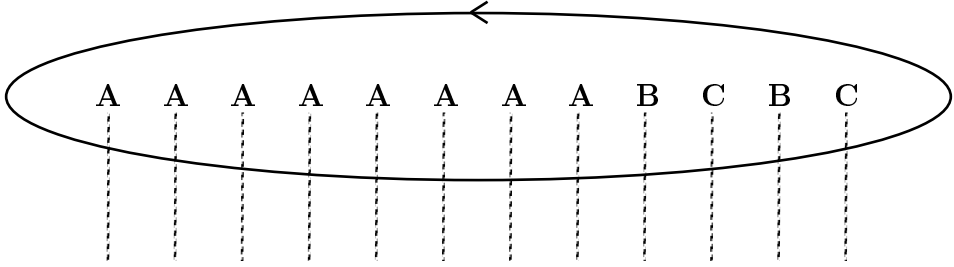}
\caption{A loop junction.}
\label{fig1}
\end{figure}

For $\wh\bE_{\bf 9}$ the junction lattice forms
an indefinite lattice with signature $(2+,10-)$.
Under the condition that the total $(p,q)$ charges (\ref{totalpq}) vanish, 
this lattice reduces to the semi-definite one isomorphic to
\beq
\bZ \bfgr{\delta}_1 \oplus \bZ \bfgr{\delta}_2 \oplus (-E_8).
\eeq
Here $\bfgr{\delta}_1$ and $\bfgr{\delta}_2$ are the null junctions. 
A choice of basis is
\beqa
&&\bfgr{\delta}_1=(0,0,0,0,0,0,0,0,-1,-1,1,1), \CR
&&\bfgr{\delta}_2=(-1,-1,-1,-1,-1,-1,-1,-1,7,5,-3,-1).
\label{null-basis}
\eeqa
The $\bfgr{\delta}_1$, $\bfgr{\delta}_2$ junctions are expressed as loops of
$(1,0)$, $(0,1)$ strings, respectively, surrounding 
the $\wh\bE_{\bf 9}$ branes (\ref{hatE9}) counterclockwise, 
see Figure \ref{fig1}. These null junctions represent two imaginary roots
of the loop algebra $\wh E_9$ \cite{DHIZ-3} and will play a crucial role 
in our study of torsions in section 4.

\subsection{Mordell-Weil lattice from junctions}

A rational elliptic surface $S$ is a special kind of
9 points blown-up of
$\bP^2$ and $b_2=\dim H_2(S)=10$.
$H_2(S)$ is a unimodular Lorentzian lattice with signature
$(1+,9-)$. 
Let $O$ and $F$ be a class of zero section and generic fiber.
Their intersections are $O \cdot O=-1$, $O \cdot F=1$ and $F \cdot F=0$.
The orthogonal complement of $\langle O,F \rangle$ in $H_2(S)$ is
isomorphic to $-E_8$.

Some elements of $H_2(S)$ appear as components of singular fibers. 
They generate the root lattice corresponding to the singularity type.
Let $T \subset E_8$ be the lattice generated by the
components of singular fibers. Theorem 10.3 in \cite{Shio} states that
the structure of the Mordell-Weil group of the rational elliptic
surface $S$ is described as
\beq
E(K) \simeq L^{\ast} \oplus (T' /T),
\label{theorem}
\eeq
where $L^{\ast}$ is the dual of $L=T^{\perp}$ (in $E_8$)
and $T'=T \otimes {\bf Q} \cap E_8$. From (\ref{theorem}) the torsion 
subgroup $E(K)_{\rm tor}$ of $E(K)$ is
read off as $E(K)_{\rm tor}\simeq T'/T$, and hence 
$E(K)/E(K)_{\rm tor}\simeq L^*$.
According to this theorem, the computation of $E(K)$
is reduced to the embedding of lattice $T$ in $E_8$.

In \cite{OS} all the possible structures of $T$ and $E(K)$ are classified.
They are listed in Table \ref{config} where $r={\rm rank}\,E(K)$.\footnote[5]{
$E(K)=A_1^*\oplus \bra 1/6 \ket$ for No.32 and $(\bZ/2\bZ)^2$ for No.70 
in \cite{OS} should read $E(K)=\Lambda_{(23)}$ and $\bZ/4\bZ$, respectively.} 
In the last column for $E(K)$, $\u1{k}$ denotes a rank one lattice 
$\bZ\,x$ with $(x,x)=k$ and
\beq
\begin{array}{ll}
\Lambda_{(12)}=\frac{1}{6}
\left(
\begin{array}{cccc}
2&1&0&-1 \CR
1&5&3&1 \CR
0&3&6&3 \CR
-1&1&3&5
\end{array}
\right), \hskip10mm  &
\Lambda_{(17)}=\frac{1}{10}
\left(
\begin{array}{ccc}
3&1&-1 \CR
1&7&3 \CR
-1&3&7
\end{array}
\right), \\
\Lambda_{(19)}=\frac{1}{12}
\left(
\begin{array}{ccc}
7&1&2 \CR
1&7&2 \CR
2&2&4
\end{array}
\right), \hskip10mm  &
\Lambda_{(23)}= \frac{1}{6}
\left(
\begin{array}{cc}
2&1 \CR
1&2
\end{array}
\right), \\
\Lambda_{(25)}=\frac{1}{7}
\left(
\begin{array}{cc}
2&1 \CR
1&4
\end{array}
\right), \hskip10mm  &
\Lambda_{(31)}=\frac{1}{15}
\left(
\begin{array}{cc}
2&1 \CR
1&8
\end{array}
\right), \\
\Lambda_{(33)}=\frac{1}{10}
\left(
\begin{array}{cc}
2&1 \CR
1&3
\end{array}
\right).
\end{array}
\label{nonCartan}
\eeq

\def\branetable{
\begin{table}
\begin{center}
\begin{tabular}{||c|c|c|c|c||} 
\hline
No. & $r$ & $T$ & branes & $E(K)$ \CR
\hline
1 & 8 &$0$ & 
$\bA^8 \bB \bC \bB \bC$ & 
$E_8$ \CR
\hline
2 & 7 & $A_1$ & 
$(\bA^2) \bA^6 \bB \bC \bB \bC$ & 
$E_7^{\ast}$ \CR
\hline
3 & 6 &$A_2$ & 
$(\bA^3) \bA^5 \bB \bC \bB \bC$ & 
$E_6^{\ast}$ \CR
4 & & ${A_1}^{\oplus 2}$ & 
$(\bA^2)^2 \bA^4 \bB \bC \bB \bC$ & 
$D_6^{\ast}$ \CR
\hline
5 & 5 & $A_3$ & 
$(\bA^4) \bA^4 \bB \bC \bB \bC$ & 
$D_5^{\ast}$ \CR
6 & & $A_2 \oplus A_1$ & 
$(\bA^3) (\bA^2) \bA^3 \bB \bC \bB \bC$ & 
$A_5^{\ast}$ \CR
7 & & ${A_1}^{\oplus 3}$ & 
$(\bA^2)^3 \bA^2 \bB \bC \bB \bC$ & 
$D_4^{\ast} \oplus A_1^{\ast}$ \CR
\hline
8 & 4 & $A_4 $ & 
$(\bA^5) \bA^3 \bB \bC \bB \bC$ & 
$A_4^{\ast}$ \CR
9 & & $D_4 $ & 
$(\bA^4 \bB \bC) \bA^4 \bB \bC$ & 
$D_4^{\ast}$ \CR
10 & & $A_3 \oplus A_1$ & 
$(\bA^4) (\bA^2) \bA^2 \bB \bC \bB \bC$ & 
$A_3^{\ast} \oplus A_1^{\ast}$ \CR
11 & & ${A_2}^{\oplus 2}$ & 
$(\bA^3)^2 \bA^2 \bB \bC \bB \bC $ & 
$A_2^{\ast \oplus 2}$ \CR
12 & & $A_2 \oplus {A_1}^{\oplus 2}$ & 
$(\bA^3) (\bA^2)^2 \bA \bB \bC \bB \bC$ & 
$\Lambda_{(12)}$ \CR
13 & & ${A_1}^{\oplus 4}$ & 
$(\bA^2)^4 \bB \bC \bB \bC$ & 
$D_4^{\ast} \oplus \bZ/2 \bZ$ \CR
14 & & ${A_1}^{\oplus 4}$ & 
$(\bA^2)^4 \bA \br{2}{-1} \br{1}{-2} \bC$ & 
$A_1^{\ast \oplus 4}$ \CR
\hline
15 & 3 & $A_5$ & 
$(\bA^6) \bA^2 \bB \bC \bB \bC$ & 
$A_2^{\ast} \oplus A_1^{\ast}$ \CR
16 & &
$D_5$ & $(\bA^5 \bB \bC) \bA^3 \bB \bC$ & 
$A_3^{\ast}$ \CR
17 & & $A_4 \oplus A_1$ & 
$(\bA^5) (\bA^2) \bA \bB \bC \bB \bC$ & 
$\Lambda_{(17)}$ \CR
18 & & $D_4 \oplus A_1$ & 
$(\bA^4 \bB \bC) (\bA^2) \bA^2 \bB \bC$ & 
$A_1^{\ast \oplus 3}$ \CR
19 & & $A_3 \oplus A_2$ & 
$(\bA^4) (\bA^3) \bA \bB \bC \bB \bC$ & 
$\Lambda_{(19)}$ \CR
20 & & ${A_2}^{\oplus 2} \oplus A_1$ & 
$(\bA^3)^2 (\bA^2) \bB \bC \bB \bC$ & 
$A_2^{\ast} \oplus \u1{1/6}$ \CR
21 & & $A_3 \oplus {A_1}^{\oplus 2}$ & 
$(\bA^4) (\bA^2)^2 \bB \bC \bB \bC$ & 
$A_3^{\ast} \oplus \bZ/2\bZ$ \CR
22 & & $A_3 \oplus {A_1}^{\oplus 2}$ & 
$(\bA^4) (\bA^2)^2 \bA \br{2}{-1} \br{1}{-2} \bC$ & 
$A_1^{\ast \oplus 2} \oplus \u1{1/4}$ \CR
23 & & $A_2 \oplus {A_1}^{\oplus 3}$ & 
$(\bA^3) (\bA^2)^3 \br{2}{-1} \br{1}{-2} \bC$ & 
$A_1^* \oplus \Lambda_{(23)}$ \CR
24 & & ${A_1}^{\oplus 5}$ & 
$(\bA^2)^2 (\bB^2)^2 (\br{0}{1}^2) \br{2}{1}^2 $ & 
$A_1^{\ast \oplus 3} \oplus \bZ/2\bZ$ \CR
\hline
25 & 2 & $A_6$ & 
$(\bA^7) \bA \bB \bC \bB \bC$ & 
$\Lambda_{(25)}$ \CR
26 & & $D_6$ & 
$(\bA^6 \bB \bC) \bA^2 \bB \bC$ & 
$A_1^{\ast \oplus 2}$ \CR
27 & & $E_6$ & 
$(\bA^5 \bB \bC^2) \br{3}{1} \bA^3$ & 
$A_2^{\ast}$ \CR
28 & & $A_5 \oplus A_1$ & $(\bA^6) (\bA^2) \bB \bC \bB \bC$ & 
$A_2^{\ast} \oplus \bZ/2\bZ$ \CR
29 & & $A_5 \oplus A_1$ & 
$(\bA^6) (\bA^2) \bA \br{2}{-1} \br{1}{-2} \bC$ & 
$A_1^{\ast} \oplus \u1{1/6}$ \CR
30 & & $D_5 \oplus A_1$ & 
$(\bA^5 \bB \bC) (\bA^2) \bA \bB \bC$ & 
$A_1^{\ast} \oplus \u1{1/4}$ \CR
\hline
\end{tabular}
\end{center}
\end{table}
\begin{table}
\begin{center}
\begin{tabular}{||c|c|c|c|c||} 
\hline
No. & $r$ & $T$ & branes & $E(K)$ \CR
\hline
31 & 2 & $A_4 \oplus A_2$ & 
$(\bA^5) (\bA^3) \bB \bC \bB \bC$ & 
$\Lambda_{(31)}$ \CR
32 & & $D_4 \oplus A_2$ & 
$(\bA^4 \bB \bC) (\bA^3) \bA \bB \bC$ & 
$\Lambda_{(23)}$ \CR
33 & & $A_4 \oplus {A_1}^{\oplus 2}$ & 
$(\bA^5) (\bA^2)^2 \br{2}{-1} \br{1}{-2} \bC$ & 
$\Lambda_{(33)}$ \CR
34 & & $D_4 \oplus {A_1}^{\oplus 2}$ & 
$(\bA^4 \bB \bC) (\bA^2)^2 \bB \bC$ & 
$A_1^{\ast \oplus 2} \oplus \bZ/2\bZ$ \CR
35 & & $A_3 \oplus A_3$ & 
$(\bA^4)^2 \bB \bC \bB \bC$ & 
$A_1^{\ast \oplus 2} \oplus \bZ/2\bZ$ \CR
36 & & $A_3 \oplus A_3$ & 
$(\bA^4)^2 \bA \br{2}{-1} \br{1}{-2} \bC$ & 
$\u1{1/4}^{\oplus 2}$ \CR
37 & & $A_3 \oplus A_2 \oplus A_1$ & 
$(\bA^4) (\bA^3) (\bB^2) \br{1}{-3} \bC \bA$ & 
$A_1^{\ast} \oplus \u1{1/12}$ \CR
38 & & $A_3 \oplus {A_1}^{\oplus 3}$ & 
$(\bA^4) (\bA^2) (\bB^2) (\br{0}{1}^2) \bB \bC$ & 
$A_1^{\ast} \oplus \u1{1/4} \oplus \bZ/2\bZ$ \CR
39 & & ${A_2}^{\oplus 3}$ & 
$(\bA^3)^3 \br{2}{-1} \br{1}{-2} \bC$ & 
$A_2^{\ast} \oplus \bZ/3\bZ$ \CR
40 & & ${A_2}^{\oplus 2} \oplus {A_1}^{\oplus 2}$ & 
$(\bA^3)^2 (\bB^2) (\br{0}{1}^2) \bB \bC$ & 
$\u1{1/6}^{\oplus 2}$ \CR
41 & & $A_2 \oplus {A_1}^{\oplus 4}$ & 
$(\bA^3) (\bB^2) (\br{0}{1}^2) (\bB^2) (\br{0}{1}^2) \bA$ & 
$\Lambda_{(23)} \oplus \bZ/2\bZ$ \CR
42 & & ${A_1}^{\oplus 6}$ & 
$(\bA^2)^2 (\bB^2)^2 (\br{0}{1}^2) (\br{2}{1}^2)$ & 
$A_1^{\ast \oplus 2} \oplus (\bZ/2\bZ)^2$ \CR
\hline
43 & 1 & $E_7$ & 
$(\bA^6 \bB \bC^2) \br{3}{1} \bA^2$ & 
$A_1^{\ast}$ \CR
44 & & $A_7$ & 
$(\bA^8) \bB \bC \bB \bC$ & 
$A_1^{\ast} \oplus \bZ/2\bZ$ \CR
45 & & $A_7$ & 
$(\bA^8) \bA \br{2}{-1} \br{1}{-2} \bC$ & 
$\u1{1/8}$ \CR
46 & & $D_7$ & 
$(\bA^7 \bB \bC) \bA \bB \bC$ & 
$\u1{1/4}$ \CR
47 & & $A_6 \oplus A_1$ & 
$(\bA^7) (\bB^2) \br{1}{-3} \bC \bA$ & 
$\u1{1/14}$ \CR
48 & & $D_6 \oplus A_1$ & 
$(\bA^6 \bB \bC) (\bA^2) \bB \bC$ & 
$A_1^{\ast} \oplus \bZ/2\bZ$ \CR
49 & & $E_6 \oplus A_1$ & 
$(\bA^2) (\bA^5 \bB \bC^2) \br{3}{1} \bA$ & 
$\u1{1/6}$ \CR
50 & & $D_5 \oplus A_2$ & 
$(\bA^5 \bB \bC) (\bA^3) \bB \bC$ & 
$\u1{1/12}$ \CR
51 & & $A_5 \oplus A_2$ & 
$(\bA^6) (\bA^3) \br{2}{-1} \br{1}{-2} \bC$ & 
$A_1^{\ast} \oplus \bZ/3\bZ$ \CR
52 & & $D_5 \oplus {A_1}^{\oplus 2}$ & 
$(\bA^5 \bB \bC) (\bB^2) (\br{0}{1}^2) \bA$ & 
$\u1{1/4} \oplus \bZ/2\bZ$ \CR
53 & & $A_5 \oplus {A_1}^{\oplus 2}$ & 
$(\bA^6) (\bB^2) (\br{0}{1}^2) \bB \bC$ & 
$\u1{1/6} \oplus \bZ/2\bZ$ \CR
54 & & $D_4 \oplus A_3$ & 
$(\bA^4 \bB \bC) (\bA^4) \bB \bC$ & 
$\u1{1/4} \oplus \bZ/2\bZ$ \CR
55 & & $A_4 \oplus A_3$ & 
$(\bA^5) (\bA^4) \br{2}{-1} \br{1}{-2} \bC$ & 
$\u1{1/20}$ \CR
56 & & $A_4 \oplus A_2 \oplus A_1$ & 
$(\bA^5) (\bA^3) (\bB^2) \br{1}{-3} \bC$ & 
$\u1{1/30}$ \CR
57 & & $D_4 \oplus {A_1}^{\oplus 3}$ & 
$(\bA^4 \bB \bC) (\bA^2) (\bB^2) (\br{0}{1}^2)$ & 
$A_1^{\ast} \oplus (\bZ/2\bZ)^2$ \CR
58 & & ${A_3}^{\oplus 2} \oplus A_1$ & 
$(\bA^4)^2 (\bB^2) \br{1}{-3} \bC$ & 
$A_1^{\ast} \oplus \bZ/4\bZ$ \CR
59 & & $A_3 \oplus A_2 \oplus {A_1}^{\oplus 2}$ & 
$(\bA^4) (\bB^3) (\br{0}{1}^2) (\br{2}{1}^2) \br{3}{1}$ & 
$\u1{1/12} \oplus \bZ/2\bZ$ \CR
60 & & $A_3 \oplus {A_1}^{\oplus 4}$ & 
$(\bA^4) (\bB^2) (\br{0}{1}^2) (\bB^2) (\br{0}{1}^2)$ & 
$\u1{1/4} \oplus (\bZ/2\bZ)^2$ \CR
61 & & ${A_2}^{\oplus 3} \oplus A_1$ & 
$(\bA^3)^2(\bB^3) (\br{1}{-2}^2) \bC$ & 
$\u1{1/6} \oplus \bZ/3\bZ$ \CR
\hline
\end{tabular}
\end{center}
\end{table}
\begin{table}
\begin{center}
\begin{tabular}{||c|c|c|c|c||} 
\hline
No. & $r$ & $T$ & branes & $E(K)$ \CR
\hline
62 & 0 & $E_8$ & 
$(\bA^7 \bB \bC^2) \br{3}{1} \bA$ & 
$0$ \CR
63 & & $A_8$ & 
$(\bA^9) \br{2}{-1} \br{1}{-2} \bC$ & 
$\bZ/3\bZ$ \CR
64 & & $D_8$ & 
$(\bA^8 \bB \bC) \bB \bC$ & 
$\bZ/2\bZ$ \CR
65 & & $E_7 \oplus A_1$ & 
$(\bA^2) (\bA^6\bB \bC^2) \br{3}{1}$ & 
$\bZ/2\bZ$ \CR
66 & & $A_5 \oplus A_2 \oplus A_1$ & 
$(\bA^6) (\bB^3) (\br{1}{-2}^2) \bC$ & 
$\bZ/6\bZ$ \CR
67 & & ${A_4}^{\oplus 2}$ & 
$(\bA^5) (\bB^5) \br{2}{-3} \bC$ & 
$\bZ/5\bZ$ \CR
68 & & ${A_2}^{\oplus 4}$ & 
$(\bA^3) (\bB^3) (\br{0}{1}^3) (\bC^3)$ & 
$(\bZ/3\bZ)^2$ \CR
69 & & $E_6 \oplus A_2$ & 
$(\bA^3) (\bA^5 \bB \bC^2) \br{3}{1} $ & 
$\bZ/3\bZ$ \CR
70 & & $A_7 \oplus A_1$ & 
$(\bA^8) (\bB^2) \br{1}{-3} \bC$ & 
$\bZ/4\bZ$ \CR
71 & & $D_6 \oplus {A_1}^{\oplus 2}$ & 
$(\bA^6 \bB \bC) (\bB^2) (\br{0}{1}^2)$ & 
$(\bZ/2\bZ)^2$ \CR
72 & & $D_5 \oplus A_3$ & 
$(\bA^5 \bB \bC) (\bB^4) \br{1}{-2}$ & 
$\bZ/4\bZ$ \CR
73 & & ${D_4}^{\oplus 2}$ & 
$(\bA^4 \bB \bC) (\bA^4 \bB \bC)$ & 
$(\bZ/2\bZ)^2$ \CR
74 & & ${(A_3 \oplus A_1)}^{\oplus 2}$ & 
$(\bA^4) (\bB^4) (\br{0}{1}^2) (\br{2}{1}^2)$ & 
$\bZ/4\bZ \oplus \bZ/2\bZ$ \CR
\hline
\end{tabular}
\end{center}
\caption{Brane configurations}
\label{config}
\end{table}
}
\branetable

To understand this result of \cite{OS} in terms of junctions, we have to 
solve the following two problems:

\noindent
(1) Find the collapsible sub-configurations of 7-branes which support
the singularity lattice $T$ as the junction lattice.

\noindent
(2) Compute the lattice $L$ using string junctions on the 7-brane
backgrounds.

The problem (1) is solved by finding a suitable rearrangement of the 
$\wh\bE_{\bf 9}$ configurations (\ref{hatE9}) with the use of the brane 
move \cite{GHZ,DHIZ-2}
\beq
{\bf X}_{\bz_i}{\bf X}_{\bz_{i+1}}={\bf X}_{\bz_i'}{\bf X}_{\bz_{i+1}'},
\label{ab-move-z}
\eeq
where
\beqa
&({\rm a})&\bz_i'=\bz_{i+1}, \quad
\bz_{i+1}'=\bz_i+(\bz_i \times \bz_{i+1})\, \bz_{i+1},  \CR
&({\rm b})&\bz_i'=\bz_{i+1}+(\bz_i \times \bz_{i+1})\, \bz_i, \quad
\bz_{i+1}'=\bz_i
\eeqa
with $\bz_i \times \bz_j=p_i q_j-p_j q_i$ for $\bz_i=(p_i,q_i)$.
Under the brane move (\ref{ab-move-z}),
the charges $Q_i$ also change as 
\beqa
&({\rm a})&Q_i'=Q_{i+1}-(\bz_i \times \bz_{i+1})Q_i,\quad Q_{i+1}'=Q_i, \CR
&({\rm b})&Q_i'=Q_{i+1},\quad Q_{i+1}'=Q_i-(\bz_i \times \bz_{i+1})Q_{i+1},
\label{ab-move-Q}
\eeqa
in such a way that 
\beq
Q_i\bz_i+Q_{i+1}\bz_{i+1}=
Q_i'\bz_i'+Q_{i+1}'\bz_{i+1}',
\eeq
thereby the total charges are kept invariant. Moreover, what is important
is that the brane moves preserve the junction lattice.
That is, under the unimodular transformation
(\ref{ab-move-Q}) we can prove the relation
\beq
-\bJ^2=\sum_{i,j} Q_i (\bs_i, \bs_j) Q_j=
\sum_{i,j} Q_i' (\bs_i', \bs_j') Q_j'=-{\bJ'}^2.
\eeq
To this end, let us rewrite $\bJ^2$ as
\beq
-\bJ^2=\sum_i Q_i \bar{Q_i},
\hskip10mm 
\bar{Q}_i=Q_i-\sum_{k<i} (\bz_k \times \bz_i) Q_k.
\eeq
Then, invariance of $\bJ^2$ follows from 
the transformation formulae for $\bar{Q}_j$ given by
\beqa
&({\rm a})&\bar{Q}_i'=\bar{Q}_{i+1},\quad
\bar{Q}_{i+1}'=\bar{Q}_i+(\bz_i \times \bz_{i+1}) \bar{Q}_{i+1}, \CR
&({\rm b})&
\bar{Q}_{i}'=\bar{Q}_{i+1}+(\bz_i \times \bz_{i+1}) \bar{Q}_{i},\quad
\bar{Q}_{i+1}'=\bar{Q}_i.
\eeqa

Now we find brane configurations for every $T$ as shown in 
Table \ref{config} where the branes put in the parenthesis are the mutually
coinciding ones corresponding to the singular fibers. We note that
the $\bA_{\bf n}$ branes are used to represent the $A_n$ singularity 
with $n=0,1,2$, though one may use ${\bf H_n}$ as well without changing 
the structure of $E(K)$.
All the configurations in Table \ref{config} can be obtained from
the $\wh\bE_9$ configuration (\ref{hatE9}) by suitable brane moves.
This means that the lattice $T$ in Table \ref{config} can be
realized as a sub-lattice of the junction lattice on $\wh\bE_9$,
hence $T \subset E_8$.

\noindent
{\bf Example}. (No.63) \\
The move from $\bA^8 \bB \bC \bB \bC$ to 
$\bA^9 \br{2}{-1} \br{1}{-2} \bC$ is given as follows:
\beqa
&&\bA^8 \bB \bC \bB \bC
=\bA^8 \bB \bC^2 \br{3}{1}
=\bA^7 \bB \br{0}{1} \bC^2 \br{3}{1} 
=\bA^7 \bB \bA^2 \br{0}{1} \br{3}{1} \CR
&& \hskip20mm =\bA^9 \br{3}{-1} \br{0}{1} \br{3}{1} 
=\bA^8 \br{2}{-1} \bC \br{4}{1} \bA \CR
&& \hskip20mm =\bA^8 \br{2}{-1} \br{1}{-2} \bC \bA
=\bA^9 \br{2}{-1} \br{1}{-2} \bC.
\eeqa

It is also possible to find corresponding curves. Some of them have already
been given in section 2, from which it is clear how to identify a curve with
an entry of Table \ref{config}. Let us enumerate more examples based 
on \cite{MNW,YY,SZ}. The massive $\wh E_n$ ($n=8,...,1$) and
$\wh{\widetilde E}_1$ curves correspond to
Nos.\,1, 2, 3, 5, 8, 15, 25, 44 and 45, respectively. 
The massless $\wh E_n$ ($n=8,...,3,1$) and $\wh{\widetilde E}_0$ curves
correspond to Nos.\,62, 65, 69, 72, 67, 66, 70 and 63, respectively.
The Seiberg-Witten curves for ${\cal N}=2$ $SU(2)$ theory 
with various flavor symmetries \cite{SW-2} can be identified
with No.64 ($N_f=0$), No.46 ($N_f=1$),
Nos.\,26, 48, 71 ($N_f=2$), Nos.\,16, 30, 50, 52, 72 ($N_f=3$) and
Nos.\,9, 18, 32, 34, 54, 57, 73 ($N_f=4$).
In each case, the number of independent mass parameters is 
given by $r={\rm rank}\,E(K)$.
For the extremal cases $r=0$, the curves and sections
were explicitly obtained earlier in \cite{Mir}.

When we consider curves in view of Table \ref{config} there is a point to
notice. As remarked above, one can use either $\bA_{\bf n}$ or $\bH_{\bf n}$
branes to describe the $A_n$ singular fibers with $n=0,1,2$. This does not
change the structure of $E(K)$, whereas it does change the explicit form of
corresponding elliptic curves, and hence we are led to different physical
interpretation. To clarify the point, let us examine the case of
$E(K)=E_8,\, E_7^*,\, E_6^*$. The 7-branes are rewritten as
\beq
(\bA^{9-N})\bA^{N-1}\bB\bC\bB\bC =(\bA^{9-N})\br{N-6}{1}\bA^{N-1}\bB\bC^2
\eeq
for $N=8,7,6$. Since the configuration $(\bA^{9-N})\br{N-6}{1}$ is equivalent 
to $(\bA^{9-N})\bC$ up to $SL(2,\bZ)$ conjugation, further coalescence can
occur
\beq
(\bA^{9-N})\br{N-6}{1} \longrightarrow (\bA^{9-N}\br{N-6}{1})=\bH_{\bf N-1},
\label{5dto4d}
\eeq
which also yields $T=A_{8-N}$. Although the singularity
structure of $T$ is identical, it is shown in \cite{YY} that the limit
(\ref{5dto4d}) gives rise to the compactification of 5D ${\cal N}=1$ 
$E_N$ theories down to 4D ${\cal N}=2$ $E_N$ theories for $N=8,7,6$. 
Thus the structure of
$E(K)$ seems not sensitive enough to distinguish these two theories.

In solving the problem (2), we recognize that the global structure of 
the surface $S$ plays a relevant role in determining 
the lattice $L$. This point is now illuminated by working out several examples.

\subsubsection{Case of Cartan type}

\begin{figure}
\hspace{1.5cm}
\epsffile{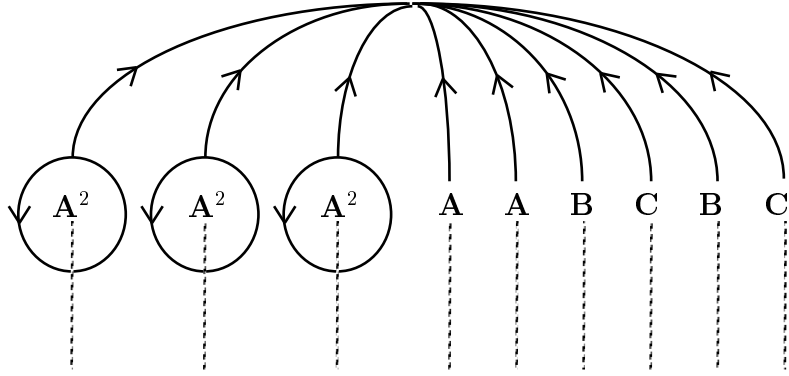}
\caption{Junctions orthogonal to $T$.}
\label{fig2}
\end{figure}

Let us first consider the case No.7 using the brane configuration 
$(\bA^2)^3 \bA^2 \bB \bC \bB \bC$.
Accordingly, we put $\bs_i={\bf a}_i$ $(1 \leq i \leq 8)$,
$\bs_9={\bf b}_1$, $\bs_{10}={\bf c}_1$, $\bs_{11}={\bf b}_2$ and
$\bs_{12}={\bf c}_2$, where ${\bf a}$, ${\bf b}$ and ${\bf c}$ denote outgoing
$(1,0)$-, $(1,-1)$- and $(1,1)$-strings attached to $\bA$, $\bB$ and $\bC$
branes, respectively. (Notice that the assignment of $\bs_j$ may be 
different from this for other cases depending
on the brane configuration.) 
Choose the root junctions generating the lattice 
$T={A_1}^{\oplus 3}$ as 
\beq
\bfgr{\alpha}_1=\bs_1-\bs_2, \hskip10mm
\bfgr{\alpha}_2=\bs_3-\bs_4, \hskip10mm
\bfgr{\alpha}_3=\bs_5-\bs_6.
\label{Tofno7}
\eeq
These junctions are supported by the collapsible branes
$\bA_1 \bA_2$, $\bA_3 \bA_4$ and $\bA_5 \bA_6$, respectively, 
and there remain 6 branes 
$\bA^2 \bB \bC \bB \bC$ of $\wh{\bE}_{\bf 9}$. From Table \ref{config} we see
that the corresponding Mordell-Weil lattice
is $E(K)=L^{\ast}=D_4^{\ast} \oplus A_1^{\ast}$. These lattices are 
canonically realized on the 7-branes $\bA^4 \bB \bC$ and $\bA^2$. Thus one 
needs apparently 8 branes, which is more than the remaining ones.
This puzzle, however, can be resolved if we consider junctions containing
strings which encircle the branes supporting $T$. 
To do so, we note that
the general form of junctions which are orthogonal to $T$ 
with $(p,q)=(0,0)$ is parameterized as
\beqa
&& \bj = Q_1(\bs_1+\bs_2)+Q_3(\bs_3+\bs_4)+Q_5(\bs_5+\bs_6)  \CR
&& \hskip7mm -(2 Q_1+2Q_3+2Q_5+Q_8+2Q_{10}+2Q_{12}) \bs_7 +Q_8 \bs_8 \CR
&& \hskip7mm 
+(Q_{10}-Q_{11}+Q_{12}) \bs_9+Q_{10}\bs_{10}+Q_{11}\bs_{11}+Q_{12}\bs_{12}.
\eeqa
Thus these junctions span the 7-dimensional lattice. This junction lattice
has elements along the null junctions (\ref{null-basis}) and the
remaining ones form a lattice of rank 5, which is expected to be isomorphic
to the lattice $D_4 \oplus A_1$.
In fact one can find the generators of $L=D_4 \oplus A_1$ as
\beqa
&&\bj_1=\bs_7+\bs_8-\bs_9-\bs_{10}, \CR
&&\bj_2=-\bs_9+\bs_{11}, \CR
&&\bj_3=\bs_5+\bs_6-\bs_9-\bs_{10}, \CR
&&\bj_4=\bs_3+\bs_4-\bs_9-\bs_{10}, \CR
&&\bj_5=\bs_7-\bs_8,
\label{D4A1basis}
\eeqa
where $\bj_i$ with $1\leq i \leq 4$ are for $D_4$ 
and $\bj_5$ is for $A_1$.
Some of the strings in these junctions are ending on the 7-branes which support
the lattice $T$ generated by (\ref{Tofno7}). It is observed, however, that
they take the form of tadpole loops of (\ref{Antad}) locally around each of
the collapsible branes $\bA_1 \bA_2$, $\bA_3 \bA_4$ and $\bA_5 \bA_6$
(see Figure \ref{fig2}),
and hence the generators (\ref{D4A1basis}) are in fact orthogonal to $T$.
It should be noted that any junction orthogonal to
the lattice $T$ in Table \ref{config} is of the form of 
(\ref{Antad})--(\ref{Entad}) locally around the coinciding
${\bf A_n}$, ${\bf D_n}$ and ${\bf E_n}$ branes which describe the singular
fibers.

In a similar manner we construct the basis junctions of $L$ for the other
cases of Cartan type intersections. The results are presented in 
Tables \ref{rgeq5}--\ref{req2and1} where $C({\cal G})$ denotes the Cartan
matrix of the Lie algebra ${\cal G}$ and the self-intersection (times $(-1)$)
of the $i$-th junction is given by the $(i,i)$ element of the intersection
matrix in the third column.

\def\meight{\left(\matrix{ 2 & -1 & 0 & 0 \cr -1 & 2 & -1 & 0 \cr 
0 & -1 & 2 & \ -1 \cr 0 & 0 & -1 & 2 \cr }\right)} 
\def\mnine{\left(\matrix{ 2 & -1 & 0 & 0 \cr -1 & 2 & -1 & -1 \cr 
0 & \ -1 & 2 & 0 \cr 0 & -1 & 0 & 2 \cr  }\right)} 
\def\mten{\left(\matrix{ 2 & 0 & 0 & 0 \cr 0 & 2 & -1 & 0 \cr 
0 & -1 & 2 & \ -1 \cr 0 & 0 & -1 & 2 \cr  }\right)} 
\def\meleven{\left(\matrix{ 2 & -1 & 0 & 0 \cr -1 & 2 & 0 & 0 \cr 
0 & 0 & 2 & \ -1 \cr 0 & 0 & -1 & 2 \cr  }\right)} 
\def\mtwelve{\left(\matrix{ 4 & -1 & 0 & 1 \cr -1 & 2 & -1 & 0 \cr 
0 & -1 & 2 & \ -1 \cr 1 & 0 & -1 & 2 \cr  }\right)} 
\def\mthirteen{\left(\matrix{ 2 & -1 & 0 & 0 \cr -1 & 2 & -1 & -1 \cr 
0 & \ -1 & 2 & 0 \cr 0 & -1 & 0 & 2 \cr  }\right)} 
\def\mfourteen{\left(\matrix{ 2 & 0 & 0 & 0 \cr 0 & 2 & 0 & 0 \cr 
0 & 0 & 2 & 0 \cr 0 & 0 & 0 & 2 \cr  }\right)} 
\def\mfifteen{\left(\matrix{ 2 & -1 & 0 \cr -1 & 2 & 0 \cr 
0 & 0 & 2 \cr  }\right)} 
\def\msixteen{\left(\matrix{ 2 & -1 & 0 \cr -1 & 2 & -1 \cr 
0 & -1 & 2 \cr  }\right)} 
\def\mseventeen{\left(\matrix{ 4 & -1 & 1 \cr 
-1 & 2 & -1 \cr 1 & -1 & 2 \cr  }\right)} 
\def\meighteen{\left(\matrix{ 2 & 0 & 0 \cr 
0 & 2 & 0 \cr 0 & 0 & 2 \cr  }\right)} 
\def\mnineteen{\left(\matrix{ 2 & 0 & -1 \cr 
0 & 2 & -1 \cr -1 & -1 & 4 \cr  }\right)} 
\def\mtwenty{\left(\matrix{ 2 & -1 & 0 \cr 
-1 & 2 & 0 \cr 0 & 0 & 6 \cr  }\right)} 
\def\mtwentyone{\left(\matrix{ 2 & -1 & 0 \cr 
-1 & 2 & -1 \cr 0 & -1 & 2 \cr  }\right)} 
\def\mtwentytwo{\left(\matrix{ 2 & 0 & 0 \cr 
0 & 2 & 0 \cr 0 & 0 & 4 \cr  }\right)} 
\def\mtwentythree{\left(\matrix{ 2 & 0 & 0 \cr 
0 & 4 & -2 \cr 0 & -2 & 4 \cr  }\right)} 
\def\mtwentyfour{\left(\matrix{ 2 & 0 & 0 \cr 
0 & 2 & 0 \cr 0 & 0 & 2 \cr  }\right)} 
\def\mtwentyfive{\left(\matrix{ 4 & -1 \cr -1 & 2 \cr  }\right)} 
\def\mtwentysix{\left(\matrix{ 2 & 0 \cr 0 & 2 \cr  }\right)} 
\def\mtwentyseven{\left(\matrix{ 2 & -1 \cr -1 & 2 \cr  }\right)} 
\def\mtwentyeight{\left(\matrix{ 2 & -1 \cr -1 & 2 \cr  }\right)} 
\def\mtwentynine{\left(\matrix{ 2 & 0 \cr 0 & 6 \cr  }\right)} 
\def\mthirty{\left(\matrix{ 2 & 0 \cr 0 & 4 \cr  }\right)} 
\def\mthirtyone{\left(\matrix{ 8 & -1 \cr -1 & 2 \cr  }\right)} 
\def\mthirtytwo{\left(\matrix{ 4 & -2 \cr -2 & 4 \cr  }\right)} 
\def\mthirtythree{\left(\matrix{ 6 & -2 \cr -2 & 4 \cr  }\right)} 
\def\mthirtyfour{\left(\matrix{ 2 & 0 \cr 0 & 2 \cr  }\right)} 
\def\mthirtyfive{\left(\matrix{ 2 & 0 \cr 0 & 2 \cr  }\right)} 
\def\mthirtysix{\left(\matrix{ 4 & 0 \cr 0 & 4 \cr  }\right)} 
\def\mthirtyseven{\left(\matrix{ 2 & 0 \cr 0 & 12 \cr  }\right)} 
\def\mthirtyeight{\left(\matrix{ 2 & 0 \cr 0 & 4 \cr  }\right)} 
\def\mthirtynine{\left(\matrix{ 2 & -1 \cr -1 & 2 \cr  }\right)} 
\def\mforty{\left(\matrix{ 6 & 0 \cr 0 & 6 \cr  }\right)} 
\def\mfortyone{\left(\matrix{ 4 & -2 \cr -2 & 4 \cr  }\right)} 
\def\mfortytwo{\left(\matrix{ 2 & 0 \cr 0 & 2 \cr  }\right)}

\begin{table}
\begin{center}
\begin{tabular}{||c|l|c||} \hline
No. & basis & $-(\bj,\bj)$ \\
\hline
{1}&$
\left\{\begin{array}{l}
\bs_{2}-\bs_{3}, \quad
\bs_{3}-\bs_{4}, \quad
\bs_{4}-\bs_{5}, \quad
\bs_{5}-\bs_{6}, \quad
\bs_{6}-\bs_{7}, \cr
\bs_{7}+\bs_{8}-\bs_{11}-\bs_{12}, \quad
\bs_{10}-\bs_{12}, \quad
\bs_{7}-\bs_{8}
\end{array}\right.$
&$C(E_8)$ \\
\hline
{2}&$
\left\{\begin{array}{l}
\bs_{3}-\bs_{4}, \quad
\bs_{4}-\bs_{5}, \quad
\bs_{5}-\bs_{6}, \quad
\bs_{6}-\bs_{7}, \cr
\bs_{7}+\bs_{8}-\bs_{11}-\bs_{12}, \quad
\bs_{10}-\bs_{12}, \quad
\bs_{7}-\bs_{8}
\end{array}\right.$
&$C(E_7)$ \\
\hline
{3}&$
\left\{\begin{array}{l}
\bs_{3}-\bs_{4}, \quad
\bs_{4}-\bs_{5}, \quad
\bs_{5}-\bs_{6}, \quad
\bs_{6}-\bs_{7}, \cr
\bs_{7}+\bs_{8}-\bs_{11}-\bs_{12}, \quad
\bs_{10}-\bs_{12}, \quad
\bs_{7}-\bs_{8}
\end{array}\right.$
&$C(E_6)$ \\
{4}&$
\left\{\begin{array}{l}
\bs_{10}-\bs_{12}, \quad
-\bs_{3}-\bs_{4}-\bs_{10}+\bs_{11}+\bs_{12}, \cr
\bs_{3}+\bs_{4}+\bs_{7}+\bs_{8}+\bs_{10}-2\bs_{11}-3\bs_{12}, \cr
\bs_{6}-\bs_{7}, \quad
\bs_{5}-\bs_{6}, \quad
\bs_{7}-\bs_{8}
\end{array}\right.$
&$C(D_6)$ \\
\hline
{5}&$
\left\{\begin{array}{l}
\bs_{10}-\bs_{12}, \quad
\bs_{7}+\bs_{8}-\bs_{9}-\bs_{10}, \cr
\bs_{6}-\bs_{7}, \quad
\bs_{5}-\bs_{6}, \quad
\bs_{7}-\bs_{8}
\end{array}\right.$
&$C(D_5)$ \\
{6}&$
\left\{\begin{array}{l}
\bs_{4}+\bs_{5}+\bs_{6}+\bs_{7}+\bs_{10}-2\bs_{11}-3\bs_{12}, \cr
-\bs_{7}+\bs_{8}, \quad
-\bs_{6}+\bs_{7}, \cr
-\bs_{7}-\bs_{8}-\bs_{10}+\bs_{10}+2 \bs_{12}, \quad
\bs_{10}-\bs_{12}
\end{array}\right.$
&$C(A_5)$ \\
{7}&$
\left\{\begin{array}{l}
\bs_{7}+\bs_{8}-\bs_{9}-\bs_{10}, \cr
-\bs_{9}+\bs_{11}, \quad
\bs_{5}+\bs_{6}-\bs_{9}-\bs_{10}, \cr
\bs_{3}+\bs_{4}-\bs_{9}-\bs_{10}, \cr
\bs_{7}-\bs_{8}
\end{array}\right.$
&$
\left(
\begin{array}{ccccc}
2&-1&0&0&0 \\
-1&2&-1&-1&0 \\
0&-1&2&0&0\\
0&-1&0&2&0\\
0&0&0&0&2
\end{array}
\right)
$ \\
\hline
\end{tabular}
\end{center}
\caption{The basis junctions for $r \geq 5$.}
\label{rgeq5}
\end{table}

\begin{table}
\begin{center}
\begin{tabular}{||c|l|c||} \hline
No. & basis & $-(\bj,\bj)$ \\
\hline
{8}&$
\left\{\begin{array}{l}
\bs_{10}-\bs_{12}, \cr
\bs_{7}+\bs_{8}-\bs_{11}-\bs_{12}, \cr
\bs_{6}-\bs_{7}, \cr
\bs_{7}-\bs_{8}
\end{array} \right.$
&$\meight$\\
{9}&$
\left\{\begin{array}{l}-\bs_{7}+\bs_{8}, \cr 
-\bs_{8}+\bs_{9}, \cr -\bs_{9}+\bs_{10}, \cr 
-\bs_{9}-\bs_{10}+\bs_{11}+\bs_{12} 
\end{array}\right.$&
$\mnine$ \\
{10}&$
\left\{\begin{array}{l}
-\bs_{7}+\bs_{8}, \cr 
-\bs_{7}-\bs_{8}+\bs_{11}+\bs_{12}, \cr 
-\bs_{10}+\bs_{12}, \cr 
-\bs_{5}-\bs_{6}+\bs_{11}+\bs_{12} 
\end{array}\right.$&
$\mten$ \\
{11}&$
\left\{\begin{array}{l}
\bs_{4}+\bs_{5}+\bs_{6}+\bs_{7}+\bs_{10}-2 \bs_{11}-3 \bs_{12}, \cr 
-\bs_{7}+\bs_{8}, \cr -\bs_{10}+\bs_{12}, \cr 
-\bs_{7}-\bs_{8}+\bs_{11}+\bs_{12} 
\end{array}\right.$&
$\meleven$ \\
{12}&$
\left\{\begin{array}{l}
\bs_{4}+\bs_{5}+2 \bs_{8}+\bs_{10}-2 \bs_{11}-3 \bs_{12}, \cr 
\bs_{6}+\bs_{7}-\bs_{11}-\bs_{12}, \cr 
\bs_{10}-\bs_{12}, \cr 
\bs_{4}+\bs_{5}-\bs_{11}-\bs_{12} 
\end{array}\right.$&
$\mtwelve$ \\
{13}&$
\left\{\begin{array}{l}-\bs_{7}-\bs_{8}+\bs_{11}+\bs_{12}, \cr 
-\bs_{10}+\bs_{12}, \cr 
-\bs_{3}-\bs_{4}+\bs_{11}+\bs_{12}, \cr 
-\bs_{5}-\bs_{6}+\bs_{11}+\bs_{12} 
\end{array}\right.$&
$\mthirteen$ \\
{14}&$
\left\{\begin{array}{l}
\bs_{3}+\bs_{4}+\bs_{9}-\bs_{11}-2 \bs_{12}, \cr 
\bs_{5}+\bs_{6}+\bs_{9}-\bs_{11}-2 \bs_{12}, \cr 
\bs_{7}+\bs_{8}+\bs_{9}-\bs_{11}-2 \bs_{12}, \cr
\bs_{3}+\bs_{4}+\bs_{5}+\bs_{6}+\bs_{7}+\bs_{8}-2 \bs_{11}-4 \bs_{12} 
\end{array}\right.$&
$\mfourteen$ \\
\hline
\end{tabular}
\end{center}
\caption{The basis junctions for $r=4$.}
\label{req4}
\end{table}

\begin{table}
\begin{center}
\begin{tabular}{||c|l|c||} \hline
No. & basis & $-(\bj,\bj)$ \\
\hline
{15}&$
\left\{\begin{array}{l}
\bs_{7}+\bs_{8}+\bs_{10}-\bs_{11}-2 \bs_{12}, \cr 
-\bs_{10}+\bs_{12}, \cr \bs_{7}-\bs_{8} 
\end{array}\right.$&
$\mfifteen$ \\
{16}&$
\left\{\begin{array}{l}
\bs_{8}-\bs_{9}, \cr -\bs_{8}+\bs_{10}, \cr 
\bs_{8}+\bs_{9}-\bs_{11}-\bs_{12} 
\end{array}\right.$&
$\msixteen$ \\
{17}&$
\left\{\begin{array}{l}
2 \bs_{8}+\bs_{10}-\bs_{11}-2 \bs_{12}, \cr 
\bs_{6}+\bs_{7}-\bs_{11}-\bs_{12}, \cr \bs_{10}-\bs_{12} 
\end{array}\right.$&
$\mseventeen$ \\
{18}&$
\left\{\begin{array}{l}
\bs_{9}-\bs_{10}, \cr \bs_{7}+\bs_{8}-\bs_{11}-\bs_{12}, \cr 
\bs_{9}+\bs_{10}-\bs_{11}-\bs_{12} 
\end{array}\right.$&
$\meighteen$ \\
{19}&$
\left\{\begin{array}{l}
\bs_{5}+\bs_{6}+\bs_{7}+\bs_{8}+\bs_{10}-2 \bs_{11}-3 \bs_{12}, \cr 
\bs_{10}-\bs_{12}, \cr -2 \bs_{8}-\bs_{10}+\bs_{11}+2 \bs_{12} 
\end{array}\right.$&
$\mnineteen$ \\
{20}&$
\left\{\begin{array}{l}
\bs_{10}-\bs_{12}, \cr \bs_{7}+\bs_{8}-\bs_{11}-\bs_{12}, \cr 
2 \bs_{4}+2 \bs_{5}+2 \bs_{6}+\bs_{7}+\bs_{8}+2 \bs_{10}-4 \bs_{11}-6 \bs_{12} 
\end{array}\right.$&
$\mtwenty$ \\
{21}&$
\left\{\begin{array}{l}
\bs_{10}-\bs_{12}, \cr \bs_{5}+\bs_{6}-\bs_{11}-\bs_{12}, \cr 
-\bs_{5}-\bs_{6}-\bs_{7}-\bs_{8}-\bs_{10}+2 \bs_{11}+3 \bs_{12} 
\end{array}\right.$&
$\mtwentyone$ \\
{22}&$
\left\{\begin{array}{l}
\bs_{5}+\bs_{6}+\bs_{9}-\bs_{11}-2 \bs_{12}, \cr 
\bs_{7}+\bs_{8}+\bs_{9}-\bs_{11}-2 \bs_{12}, \cr 
\bs_{5}+\bs_{6}+\bs_{7}+\bs_{8}-\bs_{9}-\bs_{11}-2 \bs_{12} 
\end{array}\right.$&
$\mtwentytwo$ \\
{23}&$
\left\{\begin{array}{l}
\bs_{4}+\bs_{5}+\bs_{6}+\bs_{7}+\bs_{8}+\bs_{9}-2 \bs_{11}-4 \bs_{12}, \cr 
-\bs_{4}-\bs_{5}+\bs_{6}+\bs_{7}, \cr \bs_{4}+\bs_{5}-\bs_{8}-\bs_{9} 
\end{array}\right.$&
$\mtwentythree$ \\
{24}&$
\left\{\begin{array}{l}
\bs_{3}+\bs_{4}-\bs_{7}-\bs_{8}-\bs_{9}-\bs_{10}, \cr 
\bs_{3}+\bs_{4}+\bs_{7}+\bs_{8}+2 \bs_{9}+2 \bs_{10}-\bs_{11}-\bs_{12}, \cr 
\bs_{11}-\bs_{12} 
\end{array}\right.$&
$\mtwentyfour$ \\
\hline
\end{tabular}
\end{center}
\caption{The basis junctions for $r=3$.}
\label{req3}
\end{table}

\begin{table}
\begin{center}
\begin{tabular}{||c|l|c||} \hline
No. & basis & $-(\bj,\bj)$ \\
\hline
{25}&$
\left\{\begin{array}{l}
-2 \bs_{8}+\bs_{11}+\bs_{12}, \cr -\bs_{10}+\bs_{12} 
\end{array}\right.$&
$\mtwentyfive$ \\
{26}&$
\left\{\begin{array}{l}
-\bs_{9}-\bs_{10}+\bs_{11}+\bs_{12}, \cr -\bs_{9}+\bs_{10} 
\end{array}\right.$&
$\mtwentysix$ \\
{27}&$
\left\{\begin{array}{l}
-\bs_{10}+\bs_{11}, \cr -\bs_{11}+\bs_{12} 
\end{array}\right.$&
$\mtwentyseven$ \\
{28}&$
\left\{\begin{array}{l}
-\bs_{10}+\bs_{12}, \cr -\bs_{7}-\bs_{8}+\bs_{11}+\bs_{12} 
\end{array}\right.$&
$\mtwentyeight$ \\
{29}&$
\left\{\begin{array}{l}
-\bs_{7}-\bs_{8}-\bs_{9}+\bs_{11}+2 \bs_{12}, \cr 
\bs_{7}+\bs_{8}-2 \bs_{9} 
\end{array}\right.$&
$\mtwentynine$ \\
{30}&$\left\{\begin{array}{l}
-\bs_{8}-\bs_{9}+\bs_{11}+\bs_{12}, \cr 
-2 \bs_{10}+\bs_{11}+\bs_{12} 
\end{array}\right.$&
$\mthirty$ \\
{31}&$
\left\{\begin{array}{l}
2 \bs_{6}+2 \bs_{7}+2 \bs_{8}+2 \bs_{10}-3 \bs_{11}-5 \bs_{12}, \cr 
-\bs_{10}+\bs_{12} 
\end{array}\right.$&
$\mthirtyone$ \\
{32}&$
\left\{\begin{array}{l}
-\bs_{7}-\bs_{8}-\bs_{9}+\bs_{10}+\bs_{11}+\bs_{12}, \cr 
-2 \bs_{10}+\bs_{11}+\bs_{12} 
\end{array}\right.$&
$\mthirtytwo$ \\
{33}&$
\left\{\begin{array}{l}
-\bs_{6}-\bs_{7}-2 \bs_{8}-2 \bs_{9}+2 \bs_{11}+4 \bs_{12}, \cr 
-\bs_{6}-\bs_{7}+\bs_{8}+\bs_{9} 
\end{array}\right.$&
$\mthirtythree$ \\
{34}&$
\left\{\begin{array}{l}
\bs_{9}+\bs_{10}-\bs_{11}-\bs_{12}, \cr \bs_{7}+\bs_{8}-\bs_{11}-\bs_{12} 
\end{array}\right.$&
$\mthirtyfour$ \\
{35}&$
\left\{\begin{array}{l}
-\bs_{5}-\bs_{6}-\bs_{7}-\bs_{8}-\bs_{10}+2 \bs_{11}+3 \bs_{12}, \cr 
-\bs_{10}+\bs_{12} 
\end{array}\right.$&
$\mthirtyfive$ \\
{36}&$
\left\{\begin{array}{l}
-\bs_{5}-\bs_{6}-\bs_{7}-\bs_{8}+\bs_{9}+\bs_{11}+2 \bs_{12}, \cr 
\bs_{5}+\bs_{6}+\bs_{7}+\bs_{8}+2 \bs_{9}-2 \bs_{11}-4 \bs_{12} 
\end{array}\right.$&
$\mthirtysix$ \\
{37}&$
\left\{\begin{array}{l}
-\bs_{5}-\bs_{6}-\bs_{7}+\bs_{10}+3 \bs_{11}-\bs_{12}, \cr 
\bs_{5}+\bs_{6}+\bs_{7}-3 \bs_{12} 
\end{array}\right.$&
$\mthirtyseven$ \\
\hline
\end{tabular}
\end{center}
\caption{The basis junctions for $r=2$ (continued on Table 7).}
\label{req2}
\end{table}

\begin{table}
\begin{center}
\begin{tabular}{||c|l|c||} \hline
No. & basis & $-(\bj,\bj)$ \\
\hline
{38}&$
\left\{\begin{array}{l}
-\bs_{5}-\bs_{6}+\bs_{11}+\bs_{12}, \cr 
-\bs_{5}-\bs_{6}-\bs_{9}-\bs_{10}+2 \bs_{12} 
\end{array}\right.$&
$\mthirtyeight$ \\
{39}&$
\left\{\begin{array}{l}
\bs_{4}+\bs_{5}+\bs_{6}-\bs_{11}-2 \bs_{12}, \cr 
\bs_{7}+\bs_{8}+\bs_{9}-\bs_{11}-2 \bs_{12} 
\end{array}\right.$&
$\mthirtynine$ \\
{40}&$
\left\{\begin{array}{l}
-2 \bs_{4}-2 \bs_{5}-2 \bs_{6}-\bs_{9}-\bs_{10}+2 \bs_{11}+4 \bs_{12}, \cr 
-\bs_{9}-\bs_{10}-\bs_{11}+\bs_{12} 
\end{array}\right.$&
$\mforty$ \\
{41}&$
\left\{\begin{array}{l}
\bs_{8}+\bs_{9}+\bs_{10}+\bs_{11}-2 \bs_{12}, \cr 
-\bs_{6}-\bs_{7}+\bs_{10}+\bs_{11} 
\end{array}\right.$&
$\mfortyone$ \\
{42}&$
\left\{\begin{array}{l}
\bs_{3}+\bs_{4}+\bs_{7}+\bs_{8}+2 \bs_{9}+2 \bs_{10}-\bs_{11}-\bs_{12}, \cr 
\bs_{3}+\bs_{4}-\bs_{7}-\bs_{8}-\bs_{9}-\bs_{10} 
\end{array}\right.$&
$\mfortytwo$ \\
\hline
43 & \hskip4mm $\bs_{11}-\bs_{12}$& 2 \cr
44 & \hskip4mm $\bs_{10}-\bs_{12}$& 2 \cr
45 & \hskip4mm $3\bs_{9}-\bs_{11}-2\bs_{12}$ & 8 \cr
46 & \hskip4mm $2\bs_{10}-\bs_{11}-\bs_{12}$ & 4 \cr
47 & \hskip4mm $\bs_{10}+3\bs_{11}-4\bs_{12}$ & 14 \cr
48 & \hskip4mm $\bs_{9}+\bs_{10}-\bs_{11}-\bs_{12}$ & 2 \cr
49 & \hskip4mm $\bs_{3}+\cdots
+\bs_{7}+2\bs_{9}+2\bs_{10}-4\bs_{11}+3\bs_{12}$ & 6 \cr
50 & \hskip4mm $2\bs_{8}+2\bs_{9}+2\bs_{10}-3\bs_{11}-3\bs_{12}$ & 12 \cr
51 & \hskip4mm $\bs_{7}+\bs_{8}+\bs_{9}-\bs_{11}-2 \bs_{12}$ & 2 \cr
52 & \hskip4mm $\bs_{8}+\bs_{9}+\bs_{10}-\bs_{11}-2 \bs_{12}$ & 4 \cr
53 & \hskip4mm $\bs_{9}+\bs_{10}+\bs_{11}-\bs_{12} $ & 6 \cr
54 & \hskip4mm $\bs_{7}+\bs_{8}+\bs_{9}+\bs_{10}-2 \bs_{11}-2 \bs_{12}$ & 4 \cr
55 & \hskip4mm $3\bs_{6}+3\bs_{7}+3\bs_{8}-4\bs_{11}-8\bs_{12}$ & 20 \cr
56 & \hskip4mm $4\bs_{6}+4\bs_{7}+4\bs_{8}-3\bs_{11}-9\bs_{12}$ & 30 \cr
57 & \hskip4mm $\bs_{7}+\bs_{8}-\bs_{9}-\bs_{10}-\bs_{11}-\bs_{12}$ & 2 \cr
58 & \hskip4mm $\bs_{5}+\bs_{6}+\bs_{7}+\bs_{8}-\bs_{11}-3\bs_{12}$ & 2 \cr
59 & \hskip4mm $\bs_{8}+\bs_{9}-3\bs_{10}-3\bs_{11}+4\bs_{12}$ & 12 \cr
60 & \hskip4mm $\bs_{7}+\bs_{8}-\bs_{11}-\bs_{12}$ & 4 \cr
61 & \hskip4mm $2\bs_{4}+2\bs_{5}+2\bs_{6}-\bs_{10}-\bs_{11}-4\bs_{12}$ & 6 \cr
\hline
\end{tabular}
\end{center}
\caption{The basis junctions for $r=2$ (continued) and $1$.}
\label{req2and1}
\end{table}

\subsubsection{Case of non-Cartan type}

In Table \ref{config} there exists a class of $E(K)$ which are characterized
by the intersection matrices of non-Cartan type (\ref{nonCartan}).
It is intriguing that we can indeed derive these 
intersection matrices from the junctions.
Some examples are given as follows:

\noindent {\bf Example}. (No.12) \\
We have the 7-branes $(\bA^3)(\bA^2)^2\bA\bB\bC\bB\bC$.
The junctions orthogonal to $T=A_2\oplus A_1^{\oplus 2}$ are
\beqa
&&\bj_1=\bs_{4}+\bs_{5}+2\bs_{8}+\bs_{10}-2\bs_{11}-3\bs_{12},\CR
&&\bj_2=\bs_{6}+\bs_{7}-\bs_{11}-\bs_{12}, \CR
&&\bj_3=\bs_{10}-\bs_{12}, \CR
&&\bj_4=\bs_{4}+\bs_{5}-\bs_{11}-\bs_{12}.
\eeqa
These consist of the basis junctions of $L$.
The intersection matrix turns out to be
\beq
-(\bj_i,\bj_j)=
\left(
\begin{array}{cccc}
4&-1&0&1 \CR
-1&2&-1&0 \CR
0&-1&2&-1 \CR
1&0&-1&2
\end{array}
\right).
\eeq
The inverse of this yields the matrix $\Lambda_{(12)}$ of (\ref{nonCartan}).

\noindent {\bf Example}. (No.17)\\
We have the 7-branes $(\bA^5)(\bA^2)\bA\bB\bC\bB\bC$.
The junctions orthogonal to $T=A_4\oplus A_1$ are 
\beqa
&&\bj_1=2\bs_{8}+\bs_{10}-\bs_{11}-2\bs_{12},\CR
&&\bj_2=\bs_{6}+\bs_{7}-\bs_{11}-\bs_{12}, \CR
&&\bj_3=\bs_{10}-\bs_{12},
\eeqa
whose intersection matrix is
\beq
-(\bj_i,\bj_j)=
\left(
\begin{array}{ccc}
4&-1&1 \CR
-1&2&-1 \CR
1&-1&2
\end{array}
\right).
\eeq
The inverse of this yields the matrix $\Lambda_{(17)}$ of (\ref{nonCartan}).

\noindent {\bf Example}. (No.25)\\
We have the 7-branes $(\bA^7)\bA\bB\bC\bB\bC$.
The junctions orthogonal to $T=A_6$ are 
\beqa
&&\bj_1=2\bs_{8}-\bs_{11}-\bs_{12},\CR
&&\bj_2=\bs_{10}-\bs_{12},
\eeqa
whose intersection matrix reads
\beq
-(\bj_i,\bj_j)=
\left(
\begin{array}{cc}
4&-1 \CR
-1&2
\end{array}
\right).
\eeq
The inverse of this yields the matrix $\Lambda_{(25)}$ of (\ref{nonCartan}).

Further computations enable us to write down the basis junctions and
their intersections for all the cases of non-Cartan type (\ref{nonCartan}).
The results are shown in Tables \ref{req4}--\ref{req2and1}.

Finally, for the case of ${\rm rank}\,E(K)=1$ (Nos.\,43--61), the lattice $L$
is one dimensional and we have determined its generator. The result is
presented in Table \ref{req2and1}.
Thus, we have completed the construction of the lattice $L$ listed in
Table \ref{config} in terms of string junctions, producing the classification
Tables \ref{rgeq5}--\ref{req2and1}.

\section{Weight lattice and torsions}

\renewcommand{\theequation}{4.\arabic{equation}}\setcounter{equation}{0}

\subsection{Weak integrality of invariant charges}

The Mordell-Weil group $E(K)$ consists of the free part $E(K)/E(K)_{\rm{tor}}$
and the torsion part $E(K)_{\rm{tor}}$. According to Theorem 9.2 
in \cite{Shio} the former should be isomorphic to the weight lattice 
$L^{\ast}$ of $L=T^\perp$ in $E_8$. 

We first would like to discuss the property of $L^*$ in view of junctions.
Naively, the weight junctions $\bfgr{\omega}_i$ can be obtained as
the dual to the basis junctions $\bj_i$, {\it i.e.}
$\bfgr{\omega}_i=(C^{-1})_{ij} \bj_j$, where $C$ is the intersection
matrix. Though the invariant charges $Q_i$ of these dual basis junctions
$\bfgr{\omega}_i$ are in general not integer, we can remedy this 
{\it partially} by adding null junctions.

\noindent {\bf Example}. (No.25)\\
The basis junctions of $L$ read
\beqa
&& \bj_1=(0,0,0,0,0,0,0,-2,0,0,1,1), \CR
&& \bj_2=(0,0,0,0,0,0,0,0,0,-1,0,1),
\eeqa
for which the dual basis junctions $\bfgr{\omega}_1, \bfgr{\omega}_2$ are
obtained as
\beqa
\bfgr{\omega}_1 &=& {1\over 7}\, (2\,\bj_1+\bj_2+3\,\bfgr{\delta}_2) \CR
&=& (-{3\over 7},-{3\over 7},-{3\over 7},-{3\over 7},-{3\over 7},-{3\over 7},
-{3\over 7},-1,3,2,-1,0), \CR
\bfgr{\omega}_2 &=& {1\over 7}\, (\bj_1+4\,\bj_2+5\,\bfgr{\delta}_2) \CR
&=& (-{5\over 7},-{5\over 7},-{5\over 7},-{5\over 7},-{5\over 7},-{5\over 7},
-{5\over 7},-1,5,3,-2,0),
\eeqa
where $\bfgr{\delta}_2$ is given by (\ref{null-basis}).

\noindent {\bf Example}. (No.7)\\
We have $L=D_4 \oplus A_1$. Since $L$ is generated by (\ref{D4A1basis}),
the basis of the weight lattice is found to be
\beqa
&&\bfgr{\omega}_1=(0,0,\frac{1}{2},\frac{1}{2},\frac{1}{2},\frac{1}{2},
1,1,-3,-2,1,0), \CR
&&\bfgr{\omega}_2=(0,0,1,1,1,1,1,1,-5,-3,2,0), \CR
&&\bfgr{\omega}_3=(-\frac{1}{2},-\frac{1}{2},0,0,\frac{1}{2},\frac{1}{2},
0,0,0,0,0,0), \CR
&&\bfgr{\omega}_4=(-\frac{1}{2},-\frac{1}{2},\frac{1}{2},\frac{1}{2},0,0,
0,0,0,0,0,0), \CR
&&\bfgr{\omega}_5=(\frac{1}{2},\frac{1}{2},\frac{1}{2},\frac{1}{2},\frac{1}{2},
\frac{1}{2},1,0,-3,-2,1,0).
\eeqa

These examples show that some of the charges $Q_i$ still remain fractional.
However, the fractionality is restricted by certain condition which we call
{\it weak} integrality.
In general, a junction orthogonal to the lattice $T$
can be represented by using tadpole junctions ($\bJ_2$ in Figure \ref{parts})
that go around the collapsed branes and do not touch them directly.
Accordingly, for each of the collapsed branes, say
$(\br{p_i}{q_i}\cdots\br{p_j}{q_j})$,
we require the integrality not for {\it individual} 
charges $Q_k$ ($i\leq k \leq j$),
but for their {\it total} $(p,q)$ charges. Namely
\beq
(p,q)=Q_i (p_i,q_i)+\cdots+Q_j (p_j,q_j) \in {\bf Z}^2.
\eeq
We call this condition {\it weak integrality}.

We now wish to point out that
the junctions orthogonal to $T$ subject to this weak
integrality condition form the full Mordell-Weil lattice 
$E(K)$ rather than the lattice $L$ (called the narrow Mordell-Weil lattice
in \cite{Shio,OS}). To see this, we have first checked that the torsion free 
part of $E(K)$ is the dual lattice $L^{\ast}$ of $L$ by explicitly 
constructing the dual basis in terms of string junctions.
The dual basis junctions have been worked out for all the cases
Nos.\,1--61. They indeed satisfy the weak integrality condition.

\subsection{Torsions as fractional null junctions}

Let us next consider the torsion part $E(K)_{\rm tor}$.
By definition, a section $P \in E(K)$ is a torsion if and only if
$m P=O$ for some $m \in \bZ_{>0}$. 
Then its height pairing vanishes $(P,Q)=0$ for any 
$Q \in E(K)$ since $m (P,Q)=(O,Q)=0$. 
Hence $P$ corresponds to a null junction.
We know there are two independent null junctions $\bfgr{\delta}_1$, 
$\bfgr{\delta}_2$. By virtue of the Hanany-Witten effect, 
these null junctions can be transformed
to the canonical form with integer charges $Q_i$, see (\ref{null-basis}) 
for instance. We call such null junctions strongly integral ones.
As we remarked above,
a fractional junction which is integral only in weak sense
is also allowed for $E(K)$.
It is then shown that the weakly integral null junctions, which we refer to
as fractional loop junctions, are identified as the torsions
(modulo strongly integral null junctions).
Our idea is explained by presenting some examples explicitly.

\noindent {\bf Example}. (No.73)\\
The brane configuration consists of
two collapsed branes $(\bA^4 \bB \bC)(\bA^4 \bB \bC)$.
Each of them supports one of the $D_4$ components in $T=D_4 \oplus D_4$.
The junctions orthogonal to $T$ become null junctions of the form
\beqa
\bj&=&Q_1 (\bs_1+\bs_2+\bs_3+\bs_4)-(2 Q_1+Q_{12}) \bs_5-Q_{12} \bs_6 \CR
&&-Q_1 (\bs_7+\bs_8+\bs_9+\bs_{10})+(2 Q_1+Q_{12}) \bs_{11}+Q_{12} \bs_{12}.
\eeqa
Thus the $(p,q)$ charges exchanged between the two $D_4$ components are
$p=2(Q_1-Q_{12})$ and $q=2(Q_1+Q_{12})$.
The integrality condition on $(p,q)$ requires 
$2Q_1 \equiv 2 Q_{12} \equiv 0$.
Hence we obtain the $(\bZ/2\bZ)^2$ torsion.
The generators of the torsion junctions are 
\beqa
&& \bj =
(0,0,0,0,-\frac{1}{2},-\frac{1}{2},
0,0,0,0,\frac{1}{2},\frac{1}{2}), \CR
&& \bj' =
(-\frac{1}{2},-\frac{1}{2},-\frac{1}{2},-\frac{1}{2},
\frac{3}{2},\frac{1}{2},
\frac{1}{2},\frac{1}{2},\frac{1}{2},\frac{1}{2},
-\frac{3}{2},-\frac{1}{2})
\eeqa
which in fact obey $\bj^2=\bj'^2=0$ and
$2\,\bj =\bfgr{\delta}_1,\, 2\,\bj' =\bfgr{\delta}_2$.
They are represented as the fractional loop junctions with the charges
$(r,s)=(\half, 0)$ for $\bj$ and $(r,s)=(0,\half)$ for $\bj'$.\footnote{
We note again that the charges $Q_i$ for $\bfgr{\delta}_1, \bfgr{\delta}_2$ 
depend on the brane configurations.}

\noindent {\bf Example}. (No.66)\\
We have the 7-branes $(\bA^6) (\bB^3) (\br{1}{-2}^2) \bC$ and
the torsion group reads $\bZ/6\bZ$. This is generated by the torsion junction
\beq
\bj =
(-\frac{1}{6},-\frac{1}{6},-\frac{1}{6},-\frac{1}{6},-\frac{1}{6},
-\frac{1}{6},\frac{2}{3},\frac{2}{3},\frac{2}{3},
-\frac{1}{2},-\frac{1}{2},0)
\eeq
which obeys $\bj^2=0$ and $6\,\bj =\bfgr{\delta}_1+\bfgr{\delta}_2$.
As a fractional loop junction this carries the charges 
$(r,s)=({1\over 6},{1\over 6})$.

\noindent {\bf Example}. (No.74)\\
We have the 7-branes $(\bA^4) (\bB^4) (\br{0}{1}^2) (\br{2}{1}^2)$ and
the torsion group reads $\bZ/4\bZ \oplus \bZ/2\bZ$. This is generated by
the torsion junctions
\beqa
&& \bj =
(-\frac{1}{4},-\frac{1}{4},-\frac{1}{4},-\frac{1}{4},
\frac{3}{4},\frac{3}{4},\frac{3}{4},\frac{3}{4},
2,2,-\frac{1}{2},-\frac{1}{2}), \CR
&& \bj' =
(0,0,0,0,-\frac{1}{2},-\frac{1}{2},-\frac{1}{2},-\frac{1}{2},
-\frac{3}{2},-\frac{3}{2},\frac{1}{2},\frac{1}{2})
\eeqa
which obey $\bj^2=\bj'^2=0$ and
$4\,\bj =\bfgr{\delta}_2,\,  2\,\bj'=\bfgr{\delta}_1$.
As fractional loop junctions they carry the charges 
$(r,s)=(0,{1\over 4})$ for $\bj$ and $(r,s)=(\half,0)$ for
$\bj'$.

Following this procedure we can express the generators of all the torsion 
groups in Table \ref{config} as the fractional loop junctions with the $(r,s)$
charges. Our result is 

$\bZ/2 \bZ$ :   $(r,s)=(\frac{1}{2},\frac{1}{2})$
\ Nos.\,13, 21, 28, 34, 35, 38, 44, 48, 53, 54, 59, 64, 65.

$\bZ/2 \bZ$ :   $(r,s)=(0,\frac{1}{2})$
\ No.\,24.

$\bZ/2 \bZ$ :   $(r,s)=(\frac{1}{2},0)$
\ Nos.\,41, 52.

$\bZ/3 \bZ$ :   $(r,s)=(\frac{1}{3},\frac{1}{3})$
\ Nos.\,39, 51, 61, 63.

$\bZ/3 \bZ$ :   $(r,s)=(0,\frac{1}{3})$ 
\ No.\,69.

$\bZ/4 \bZ$ :   $(r,s)=(\frac{1}{4},\frac{1}{4})$
\ Nos.\,58, 70.

$\bZ/4 \bZ$ :   $(r,s)=(\frac{1}{4},\frac{1}{2})$
\ No.\,72.

$\bZ/5 \bZ$ :   $(r,s)=(\frac{1}{5},\frac{1}{5})$
\ No.\,67.

$\bZ/6 \bZ$ :   $(r,s)=(\frac{1}{6},\frac{1}{6})$
\ No.\,66.

$(\bZ/2 \bZ)^2$ :   $(r,s)=(\frac{1}{2},0),(0,\frac{1}{2})$
\ Nos.\,42, 57, 60, 71, 73.

$(\bZ/3 \bZ)^2$ :   $(r,s)=(\frac{1}{3},0),(0,\frac{1}{3})$
\ No.\,68.

$\bZ/4\bZ \oplus \bZ/2\bZ$ :   $(r,s)=(0,\frac{1}{4}),(\frac{1}{2},0)$
\ No.\,74.

\noindent
The result for the torsion group we have obtained from the junction
consideration is in agreement with that in \cite{OS}.

\section{Discussion}

\renewcommand{\theequation}{5.\arabic{equation}}\setcounter{equation}{0}

In this paper, we have systematically studied the structure of 
singularities, Mordell-Weil lattices and torsions of a rational 
elliptic surface by making use of the 7-brane-junction technology. Our results
are in nice agreement with Oguiso-Shioda's Main Theorem in \cite{OS}.
Consequently we found explicit correspondence between sections and junctions,
which is summarized in Table \ref{table s-j}.

Though we have restricted ourselves to the case of rational elliptic surfaces,
generalization to other elliptic surfaces is clear
and stated as follows:\footnote{
We thank M. Saito for pointing out this interpretation. See \cite{CZ}.}
For a general elliptic surface $p : S \rightarrow C$ over a curve $C$,
the lattice of strongly integral tadpole junctions can be identified with the
cohomology group $H^1(C,R^1 p_{\ast} \bZ)$ whose $H^{1,1}$
part is isomorphic to the narrow Mordell-Weil group of $S$.
(For a rational elliptic surface, all the elements in $H^1(C,R^1 p_{\ast} \bZ)$
belong to its $H^{1,1}$ part, however, it is not so in general.)
Since the sheaf $R^1 p_{\ast} \bZ$ is a local system whose fiber at $x \in C$
is $H^1(p^{-1}(x),\bZ)=\bZ \alpha \oplus \bZ \beta$,
the cohomology $H^1(C,R^1 p_{\ast} \bZ)$ can be evaluated by using group
cohomology associated with the monodromy representation
$\rho : \pi_1(C-\{{\rm singularity} \}) \rightarrow SL(2,\bZ)$.
The 7-brane-junction technology may offer an efficient way to calculate
this cohomology $H^1(C,R^1 p_{\ast} \bZ)$ and intersections on it.

\renewcommand{\arraystretch}{1.4}
\begin{table}
\begin{center}
\begin{tabular}{||c|c|c||} \hline
& section & junction  \\
\hline
\hline
$O$ & zero section & strongly integral null junctions \\
\hline
$E(K)_{\rm tor}$  & torsions  &  weakly integral null junctions \\
\hline
$L$ & narrow Mordell-Weil lattice & strongly integral tadpole junctions \\
\hline
$E(K)$ & Mordell-Weil lattice & weakly integral tadpole junctions\\ 
\hline
\end{tabular}
\caption{Sections and junctions}
\label{table s-j}
\end{center}
\end{table}

One may apply our results to gain a physical understanding of torsion groups
which play an important role to determine the gauge {\it group}
rather than the gauge {\it algebra} \cite{AM-2}. 
Since the structure of Mordell-Weil
lattices is related to the Wilson lines on the heterotic side under
F-theory/heterotic duality, it will be interesting to think of the issue from 
the heterotic string point of view.

Another application is found when we consider 
stable non-BPS states in F-theory.
It has recently been recognized that, in string theory, there exist 
solitonic states which are stable but not BPS \cite{Sen-non}.
These states are the lightest ones which carry certain conserved charges,
{\it i.e.} there exist no other BPS or non-BPS states of lower mass having
the same charges. Thus their stability is ensured by charge conservation.
In \cite{SZ} the analysis of stable non-BPS states in F-theory on $K3$
was initiated. As we will see now, the structure of
the Mordell-Weil lattice determines the 7-brane configurations supporting
non-BPS junctions which could be candidates for stable non-BPS states in a 
region of the moduli space of F-theory on $K3$.

Let us consider the region of the moduli space 
where 24 7-branes for an elliptic
$K3$ are split into two $\wh\bE_9$ \cite{DHIZ-3}. This corresponds to the
so-called stable degeneration of $K3$ \cite{FMW,AM-1}. Two $\wh\bE_9$
brane configurations are properly isolated from each other
in the sense of \cite{SZ}. Thus we may focus on string junctions
stretched on a single $\wh\bE_9$ to analyze non-BPS states.
The BPS junctions $\bJ_{BPS}$ have to obey the holomorphy condition which
is stated as $\bJ_{BPS}^2 \geq -2$
\cite{MNS,dWHIZ}. Thus junctions $\bJ$ with $\bJ^2 < -2$ are non-BPS.
Inspecting Table \ref{req2and1} we first observe that the basis junctions
of one-dimensional $L$, except for the case $L=A_1$, are all non-BPS.
As for the higher rank $E(K)$, we also have non-BPS basis junctions in 
Nos.\,12, 17, 19, 20, 22, 23, 25, 29--33, 36--38, 40 and 41. 
We note that the cases No.45 ($\widehat{\widetilde \bE}_1$),
No.46 ($\bD_1$) and No.25 $(\wt \bE_2)$ have already appeared in \cite{SZ}.
Each non-BPS basis junction, which we denote as $\bj_{u(1)}$, 
generates the $U(1)$ symmetry associated to the one-dimensional lattice.
It is clear that any junction can be written as 
\beq
\bJ =\bj^{\bot}+n\,\bj_{u(1)},
\eeq
where $n \in \bZ$ and $\bj^{\bot}$ stands for the orthogonal components 
of $\bJ$ with respect to the $U(1)$ direction.
Then the self-intersection is obtained as 
$\bJ^2={\bj^{\bot}}^2+n^2\,\bj_{u(1)}^2$. Since ${\bj^{\bot}}^2 \leq 0$ and 
$\bj_{u(1)}^2 < -2$ we have $\bJ^2 < -2$ for any $n \not= 0$.
Therefore the junctions with non-vanishing component
along $\bj_{u(1)}$ represent non-BPS states. Among these non-BPS states 
there could be stable states against decay. 
Identifying such stable states requires the detailed dynamical analysis
which is beyond the scope of this paper.

\vskip6mm\noindent
{\bf Acknowledgements}

\vskip2mm
We would like to thank M. Noumi and M. Saito for valuable discussions. 
We also wish to thank K. Oguiso for ascertaining some corrections in the
Table in \cite{OS}
and Y. Ohtake for preparing figures in this paper.
The work of SKY was supported in part by Grant-in-Aid for Scientific Research 
on Priority Area 707 ``Supersymmetry and Unified Theory of Elementary 
Particles'', Japan Ministry of Education, Science and Culture.

\newpage



\begin{thebibliography}{99}

\bibitem{YY} Y. Yamada and S.-K. Yang, 
{\it Affine 7-brane Backgrounds and Five-Dimensional $E_N$ Theories on $S^1$},
hep-th/9907134.

\bibitem{SZ} A. Sen and B. Zwiebach, 
{\it Stable Non-BPS States in F-theory}, hep-th/9907164.

\bibitem{Vafa} C. Vafa, 
{\it Evidence for F-Theory},
Nucl. Phys. {\bf B469} (1996) 403, hep-th/9602022.

\bibitem{Sen-1} A. Sen, 
{\it F-theory and Orientifolds},
Nucl. Phys. {\bf B475} (1996) 562, hep-th/9605150.

\bibitem{MV-1} D.R. Morrison and C. Vafa, 
{\it Compactifications of F-Theory on Calabi--Yau Threefolds -- I},
Nucl. Phys. {\bf B473} (1996) 74-92, hep-th/9602114; \\
{\it Compactifications of F-Theory on Calabi--Yau Threefolds -- II},
Nucl. Phys. {\bf B476} (1996) 437, hep-th/9603161.

\bibitem{DHIZ-3} O. DeWolfe, T. Hauer, A. Iqbal and B. Zwiebach,
{\it Uncovering Infinite Symmetries on {\rm [$p,q$]} 7-branes: Kac-Moody
Algebras and Beyond}, hep-th/9812209.

\bibitem{FMW} R. Friedman, J. Morgan and E. Witten,
{\it Vector Bundles and F Theory},
Commun. Math. Phys. {\bf 187} (1997) 679, hep-th/9701162.

\bibitem{AM-1} P.S. Aspinwall and D.R. Morrison,
{\it Point-like Instantons on K3 Orbifolds},
Nucl. Phys. {\bf B503} (1997) 533, hep-th/9705104.

\bibitem{Asp} P.S. Aspinwall,
{\it Aspects of the Hypermultiplet Moduli Space in String Duality},
JHEP {\bf 9804} (1998) 019, hep-th/9802194.

\bibitem{AM-2} P.S. Aspinwall and D.R. Morrison,
{\it Non-Simply-Connected Gauge Groups and Rational Points on
Elliptic Curves},
JHEP {\bf 9807} (1998) 012, hep-th/9805206.

\bibitem{Shio} T. Shioda, 
{\it On the Mordell-Weil Lattices},
Comment. Math. Univ. St. Pauli. {\bf 39} (1990) 211.

\bibitem{OS} K. Oguiso and T. Shioda,
{\it The Mordell-Weil Lattice of a Rational Elliptic Surface},
Comment. Math. Univ. St. Pauli. {\bf 40} (1991) 83.

\bibitem{ST} J.H. Silverman and J. Tate, {\it Rational Points on Elliptic
Curves}, Undergraduate Texts in Mathematics, Springer-Verlag, 1992.

\bibitem{Kod} K. Kodaira, 
{\it On Compact Analytic Surfaces II}, Ann. Math. {\bf 77} (1963) 563;\\
{\it On Compact Analytic Surfaces III},
Ann. Math. {\bf 78} (1963) 1.

\bibitem{SW-2} N. Seiberg and E. Witten,
{\it Monopoles, Duality and Chiral Symmetry Breaking in N=2
Supersymmetric QCD},
Nucl. Phys. {\bf B431} (1994) 484, hep-th/9408099.

\bibitem{MN} J.A. Minahan and D. Nemeschansky,
{\it An N=2 Superconformal Fixed Point with $E_6$ Global Symmetry},
Nucl. Phys. {\bf B482} (1996) 142, hep-th/9608047; \\
{\it Superconformal Fixed Points with $E_n$ Global Symmetry},
Nucl. Phys. {\bf B489} (1997) 24, hep-th/9610076.

\bibitem{NTY} M. Noguchi, S. Terashima and S.-K. Yang,
{\it N=2 Superconformal Field Theory with ADE Global Symmetry on a D3-brane
Probe}, hep-th/9903215, to appear in Nucl. Phys. {\bf B}.

\bibitem{MNW} J.A. Minahan, D. Nemeschansky and N.P. Warner,
{\it Investigating the BPS Spectrum of Non-Critical $E_n$ Strings},
Nucl. Phys. {\bf B508} (1997) 64, hep-th/9705237.

\bibitem{dWZ} O. DeWolfe and B. Zwiebach, 
{\it String Junctions for Arbitrary Lie Algebra Representations},
Nucl. Phys. {\bf B541} (1999) 509, hep-th/9804210.

\bibitem{DHIZ-2} O. DeWolfe, T. Hauer, A. Iqbal and B. Zwiebach,
{\it Uncovering the Symmetries on {\rm [$p,q$]} 7-branes: Beyond the Kodaira
Classification}, hep-th/9812028.

\bibitem{GHZ} M.R. Gaberdiel, T. Hauer and B. Zwiebach,
{\it Open string - string junction transitions},
Nucl. Phys. {\bf B525} (1998) 117, hep-th/9801205.

\bibitem{Mir} R. Miranda,
{\it The Moduli of Weierstrass fibrations over $\bP^1$},
Math. Ann. {\bf 255} (1981) 379.

\bibitem{CZ} D. Cox and S. Zucker,
{\it Intersection numbers of sections of elliptic surfaces},
Invent. Math. {\bf 53} (1979) 1.

\bibitem{Sen-non} A. Sen, 
{\it Stable Non-BPS States in String Theory},
JHEP {\bf 9806} (1998) 007, hep-th/9803194; \\
{\it Stable Non-BPS Bound States of BPS D-branes},
JHEP {\bf 9808} (1998) 010, hep-th/9805019.

\bibitem{MNS} A. Mikhailov, N. Nekrasov and S. Sethi,
{\it Geometric Realizations of BPS States in N=2 Theories},
Nucl. Phys. {\bf B531} (1998) 345, hep-th/9803142.

\bibitem{dWHIZ} O. DeWolfe, T. Hauer, A. Iqbal and B. Zwiebach,
{\it Constraints on the BPS Spectrum of N = 2, D = 4 Theories with
A-D-E Flavor Symmetry},
Nucl. Phys. {\bf B534} (1998) 261, hep-th/9805220.

\end{thebibliography}
\end{document}